\documentclass[aps,prb,twocolumn,superscriptaddress,floatfix]{revtex4-2}

\usepackage[utf8]{inputenc}
\usepackage[english]{babel}
\usepackage{mathtools}
\usepackage{physics}
\usepackage{braket}
\usepackage{bbm}
\usepackage{bm}
\usepackage{amsbsy}
\usepackage{amsthm}
\usepackage{amssymb}
\usepackage{amsfonts}
\usepackage{amsmath}
\usepackage{dsfont} % for symbols like the identity: \mathds{1}
\usepackage{graphicx} % for graphics
\usepackage{hhline}
\usepackage{epsfig}
\usepackage{epstopdf}
\usepackage{dsfont}
\usepackage{multibib}
\usepackage{xcolor}
\usepackage{color}
\usepackage{verbatim}
\usepackage{nomencl}
\usepackage[colorlinks]{hyperref}
\usepackage{float}
\usepackage[normalem]{ulem}
\usepackage{multirow}
\usepackage{makecell}
\newtheorem*{defi}{Definition}

\addto\captionsenglish{}

\begin{document}
	
	\title{Genuine Multipartite Correlations in a Boundary Time Crystal}

	\author{Ant\^onio C. Louren\c{c}o}
	\email[]{lourenco.antonio.c@gmail.com}
	
	\affiliation{Departamento de F\'isica, Universidade Federal de Santa Catarina, CEP 88040-900, Florian\'opolis, SC, Brazil}
	
	\author{Luis Fernando dos Prazeres}
	\affiliation{Instituto de F\'isica, Universidade Federal Fluminense, Av. Gal. Milton Tavares de Souza s/n, Gragoat\'a, 24210-346 Niter\'oi, Rio de Janeiro, Brazil}
	
	\author{Thiago O. Maciel}
	\affiliation{Instituto de Física, Federal University of Rio de Janeiro, Rio de Janeiro 21941-972, Brazil}

	\author{Fernando Iemini}
	\affiliation{Instituto de F\'isica, Universidade Federal Fluminense, Av. Gal. Milton Tavares de Souza s/n, Gragoat\'a, 24210-346 Niter\'oi, Rio de Janeiro, Brazil}
	
	\author{Eduardo I. Duzzioni}
	\affiliation{Departamento de F\'isica, Universidade Federal de Santa Catarina, CEP 88040-900, Florian\'opolis, SC, Brazil}
	
	\begin{abstract}
 In this work we study genuine multipartite 
correlations (GMC's) in a boundary time crystal (BTC).
 Boundary time crystals are nonequilibrium quantum phases of matter in contact to an environment, for which a macroscopic fraction of the many-body system breaks the time-translation symmetry. We analyze both (i) the structure (orders) of GMC's among the subsystems, as well as (ii) their build-up dynamics for an initially uncorrelated state. 
 We find that, in the thermodynamic limit (and only in such a limit), multipartite correlations of all orders grow indefinitely in time in the BTC phase, further displaying a persistent oscillatory behavior around their mean growth. The orders of the correlations show a power-law decaying hierarchy among its $k$-partitions.
  Moreover, in the long-time limit the GMC's are shown extensive with the system size, contrasting to the subextensive scaling in the non time-crystal (ferromagnetic) phase of the model.
  We also discuss the classical and quantum nature of these correlations with basis on multipartite entanglement witnesses, specifically, the analysis of the Quantum Fisher Information (QFI). Both GMC and QFI are able to capture and distinguish the different phases of the model. Our work highlights the genuine many-body properties of these peculiar non-equilibrium phases of matter.
	\end{abstract}
	
	\maketitle

	%%%%%%%%%%%%%%%%%%%%%%%%%%%%%%%%%%%%%%%%%%%%%%%%%%
	\section{Introduction}

   Time crystals are many-body interacting systems  spontaneously breaking the time translational symmetry. The idea was first proposed by Wilczek \cite{wilczek20121} suggesting its existence in the ground state of a closed many-body system. Soon after its proposal, however, works by Bruno \cite{bruno2013}, Watanabe and Oshikawa \cite{watanabe2015} showed that such forms of time crystals are impossible (at least for not too long-ranged interacting systems), proving a no-go theorem for the existence of these phases in thermal equilibrium quantum states. These results thus indicated that a proper ground for time crystal phases are under non-equilibrium conditions. Many studies were pursued along this direction, with fruitful results, showing the existence of time crystal in driven closed systems as well as in open dynamics, breaking from a continuous to a discrete time translational symmetry~\cite{Iemini2018,russomanno2017,surace2019,sacha2015,syrwid2017,Prokofev2018,else2016,khemani2016,vonkeyserlingk2016,yao2017,ho2017,huang2018,Bua2019,russomanno2020,khasseh2019,Yao2020,hurtado2020,wang2021,piccitto2021,Riera2020,lazarides2020,Lled2020,seibold2020,seetharam2021,homann2020}.
The  experimental realization of discrete time crystals was soon realized after its proposal~\cite{Choi2017,Zhang2017,kyprianidis2021,rovny2018,pal2018}, as well as the observation of interactions between two time crystals~\cite{Autti2020} and a real-space observation in magnons systems~\cite{trager2021} . More recently, the time-crystalline eigenstate order has been observed on a quantum processor \cite{mi2021observation}.  Some reviews about time crystals are available in Refs. \cite{Sacha2017,khemani2019brief,Guo2020}.

    In order to spontaneously break time translation symmetry the quantum system must support, in the thermodynamic limit (and only in such a limit), long-range order in time, featuring therefore rigid and persistent oscillations~\cite{russomanno2017}. 
    A particular form of time crystals was proposed in Ref.~\cite{Iemini2018} for quantum systems 
    in contact to an environment, in which only macroscopic fraction of the many-body system breaks time translation symmetry, thus dubbed as boundary time crystal (BTC). Recently the study of BTC's in extended $d$-level collective systems \cite{prazeres2021} have shown a rich dynamical phenomenology. 
    In such BTC's the system was shown to break a continuous time translation symmetry, in which it self-organizes oscillating in a persistent way in the thermodynamic limit. The persistent dynamics can be observed through a local order parameter, as its macroscopic magnetization. 
    The characteristics of BTC's, however, may not be restricted to local order parameters. Quantum fluctuations, for example, can play an important role in their characterization showing an effective non-Markovian dynamics \cite{carollo2021}.
    The structure of (classical and quantum) correlations may also hinder valuable information about such peculiar phases of matter, unraveiling its genuine many-body properties. This is a subject we shall explore in this manuscript.

 In this work we study genuine multipartite correlations (GMC's) in a boundary time crystal phase. 
 We analyze both (i) the structure (orders) of GMC's  among the subsystems in the non-equilibrium steady state of the system, as well as (ii) their build-up during the dynamics of an initially uncorrelated state. 
 \begin{table*}[t]
    \caption{Main results of the paper.}
    \label{tab:table}
    \begin{tabular}{l|l|l|l}
    \hline
        & BTC phase & Ferromagnetic phase &                    \\ \hline
    GMC &  Extensive with the system size   &  \begin{tabular}{cc} Subextensive with the system size \\ (finite in the $N\rightarrow \infty$) \end{tabular}  & \multirow{2}{*}{NESS}     \\ \cline{1-3}
    QFI &  \begin{tabular}{cc} Subextensive with the system size\\ (do not witness entanglement) \end{tabular}   &  \begin{tabular}{cc} Extensive with the system size \\ (witness entanglement) \end{tabular}  &                           \\ \hline
    GMC &  Persistent oscillations around a mean algebraic growth   &  Exponential decay towards a constant value  & \multirow{2}{*}{DYNAMICS} \\ \cline{1-3}
    QFI & Persistent oscillations around a mean algebraic decay   & Exponential decay towards a constant value   &                           \\ \hline
    \end{tabular}
    \end{table*}
These correlations are shown to grown indefinitely in time, showing a persistent oscillatory behavior (around the mean growth) in the thermodynamic limit - and only in this limit.
 The system we analyze is an open quantum system of spins $1/2$ interacting collectively with a common environment~\cite{Iemini2018}. The quantifier for the GMC's we employ in our analysis was proposed by Girolami et al. \cite{girolami2017}.  Recently this measure was applied to understand the collective behavior in the Dicke superradiance~\cite{calegari2020} and in the quantum phase transition of the Lipkin-Meshkov-Glick model~\cite{lourenco2020}. In addition to the GMC's, we also study the behavior of the Quantum Fisher Information (QFI) in the system, which works as a witness for multipartite \textit{quantum (entanglement)} correlations \cite{pezze2009,toth2012,hyllus2012,pezze2018}. The QFI captures the oscillatory characteristics of the BTC dynamics, but it is not able to fully discriminate the quantum nature of their correlations. We discuss these results with connection to the purity and coherence of the NESS.
  %\sout{ Furthermore, through calculus of quantum Fisher information(QFI), we answer if the BTC is entangled along its time evolution. By analyzing the non-equilibrium steady state (NESS) we verify the amount of quantum and classical correlations that remains for long time dynamics through the GMC measure. A deeper analysis of these correlations show that the NESS is also entangled near of phase transition, even though without knowing the amount of genuine multipartite entanglement.} 

    The manuscript is organized as follows.  In Sec. \ref{sec:BTC} we present the model under study. The GMC's and QFI measures are introduced in Sec.~\ref{sec:mat}. We study in Sec. \ref{sec.ness} the properties of these correlations in the NESS of the model, and their dynamics are shown in Sec.~\ref{sec:res}. The conclusions are presented in Sec.~\ref{sec:conclusion}.

	%%%%%%%%%%%%%%%%%%%%%%%%%%%%%%%
	\section{Boundary Time Crystal}
	%%%%%%%%%%%%%%%%%%%%%%%%%%%%%%%
	\label{sec:BTC}
	
	 In this section we present the model studied in the manuscript supporting a BTC~\cite{Iemini2018}. BTC's occur at the boundary of the system, with a macroscopic fraction of the system breaking the continuous time translation symmetry, while the bulk remaining invariant in time. In a general form, the Hamiltonian of the whole system can be described as $\hat{H}=\hat{H}_B+\hat{H}_b+\hat{V}$, where $\hat{H}_B$ and $\hat{H}_b$ are the Hamiltonian of the bulk and boundary, respectively, and $\hat{V}$ is the interaction term. The whole system evolves according to the Schrödinger equation $\ket{\psi(t)}=e^{-i\hat{H}t}\ket{\psi(0)}$, where we have set $\hbar=1$, while the state of the boundary is obtained by tracing out the bulk $\hat{\rho}_b=\tr_B{\ket{\psi(t)}\bra{\psi(t)}}$. Within a Markovian approximation the dynamics of the boundary system in the interaction picture is governed by the following master equation,
	\begin{equation}
	    \label{eq:evo}
	    \frac{d}{dt}\hat{\rho}_b=\mathcal{\hat{L}}[\hat{\rho}_b],
	\end{equation}
	where $\mathcal{\hat{L}}$ is the Lindbladian super operator, a complete positive and trace preserving map. A characteristic of a BTC is the existence of an order parameter $\hat{O}_b$ for the subsystem at the boundary with $\lim_{N_b,N_B \rightarrow \infty}\Tr[\hat{O}_b\hat{\rho}_b]=f(t)$, with $f(t)$ being a time periodic function and $N_b$ ($N_B$) the number of spins, or degrees of freedom, at the boundary (in the bulk).
	
	 The physical model of the BTC studied here describes the cooperative emission of two-level systems \cite{hannukainen2018,drummond1978,puri1979,schneider2002,walls1978,walls1980}, with Lindbladian given by 
	\begin{equation}
	\label{eq:evoequation}
	\frac{d}{dt}\hat{\rho}_b=i\omega_0\left[\hat{\rho}_b, \hat{S}_x\right]+\frac{\gamma}{S}\left(\hat{S}_-\hat{\rho}_b \hat{S}_+ - \frac{1}{2}\{ \hat{S}_+\hat{S}_-,\hat{\rho}_b \} \right),
	\end{equation}
	with $\omega_0$ being the intensity of the external field, $\gamma$ is the effective decay rate, $\hat{S}_{\alpha}=\sum_{i=1}^{N}\hat{\sigma}_{\alpha}^{i}$ are collective spin operators, for which $\hat{\sigma}_{\alpha}^{k}$ are the Pauli matrices with $\alpha=x,y,z$, $S=N_b/2=N/2$ is the total spin, and $\hat{S}_{\pm}= \hat{S}_{x} \pm i\hat{S}_{y}$ are collective ladder operators of lowering and raising, respectively.  
	While for $\omega_0 < \gamma $ the model shows a trivial (time-independent) ferromagnetic steady state, in the case  $\omega_0 > \gamma $ one observes the emergence of a BTC. The oscillating frequency of the BTC is an incommensurate of the coupling costants $\omega_0/\gamma$,  thus featuring a continuous time translation symmetry breaking~\cite{Iemini2018}. We set up $\gamma=1.0$, so it will be omitted from now on.
	
	Due to the collective nature of the interactions the system conserves the total angular momentum. We work in the subspace with maximal angular momentum, described by the symmetric Dicke states $\ket{N,n_{e}}$
	\begin{equation}
	\label{eq: dicke states}
	\ket{N,n_e}=\frac{1}{\sqrt{\binom{N}{n_e}}}\sum_{i}\mathcal{P}_{i}\left(\ket{\downarrow}^{\otimes (N-n_{e})}\otimes\ket{\uparrow}^{\otimes n_{e}}\right)
	\end{equation}
	 where $n_{e}$ is the number of excited (up) spins and the sum is taken over all possible permutations of $n_e$, described by the permutation operator $\mathcal{P}_{i}$, and $\binom{N}{n_e}$ is the binomial coefficient required to normalize the Dicke state. The Dicke states are totally symmetric by permutation of their spins, a property that will be useful in order to evaluate the GMC's.
	
	 In our studies we shall compute the GMC's for different $k$-partite partitions, studying both its dynamics as well as their non-equilibrium steady states. We evolve the master equation from an initial uncorrelated pure state, specifically the ground state of $\hat{H}_b=\omega_0\hat{S}_x$, $|\psi(0) \rangle = |-\rangle ^{\otimes N}$, towards mixed combinations of Dicke states. To evolve the system density matrix we use the Runge-Kutta method of fourth order in order to solve numerically the differential master equation. The non-equilibrium steady states, reached in the asymptotic times ($t\rightarrow \infty$), are also obtained both from a direct numerical diagonalization of the Lindbladian or analytically (see Appendix \ref{appendix}). 
	 
	 %\textcolor{red}{Qual o estado inicial do sistema?}

	%%%%%%%%%%%%%%%%%%%%%%%%%%%%%%%%%%%%%%%%%%%%%%%%%%%%%%%
	\section{Measures of Genuine multipartite correlations}
	%%%%%%%%%%%%%%%%%%%%%%%%%%%%%%%%%%%%%%%%%%%%%%%%%%%%%%%
	\label{sec:mat}
	
	We introduce in this section the GMC's \cite{girolami2017} and Quantum Fisher Information (QFI) \cite{braunstein1994,BRAUNSTEIN1996,hyllus2010not}. The former can be viewed as the information that is encoded in the density matrix of a system of $N$ spins $\hat \rho_N$ that is missing for an observer that has access only to parts of the system. For instance, for an observer that has access only to the state of individual spins, it is not possible to know how spins are correlated with each other. Therefore, all information related to correlations involving two or more spins is missing for this observer. To quantify these correlations in a system, we first define the set of uncorrelated states. 
	\begin{defi}[$k$-partite genuine product states\footnote{$k$-partite here means the that there are at most $k$ subsystems inside a partition.}]
		 The set of states that have up to $k$ subsystems is defined as,
		\begin{equation}
		P_{k}\coloneqq \left \{\sigma_{N}=\bigotimes_{j=1}^{m}\sigma_{k_{j}}, \sum^{m}_{j=1}k_{j}=N, k=\max\{k_{j}\}\right \},
		\end{equation}
		where $\sigma_{k_{j}}$ is a subsystem of $k_{j}$ spins. This set contains all the sets $P_{k'}$ with $k'<k$, such that  $P_{1}\subset P_{2}...\subset P_{N-1}\subset P_{N}$.
	\end{defi}

	The multipartite correlations of order higher than $k$, denoted by  $I^{k\rightarrow N}$, is computed by the smallest distance of the state $\hat \rho_N$ to states in the set $P_k$. Despite the relative entropy being a pseudo-distance, it will be used in order to simplify the calculations, leading to
	\begin{equation}\label{eq: correlation ktoN}
	I^{k\rightarrow N}(\rho_{N}) = \min_{\sigma_N \in P_{k}}S(\rho_{N}|| \sigma_N),
	\end{equation}
	with the minimization being taken over all product states $\sigma_N = \bigotimes^{m}_{i=1} \sigma_{k_{i}} \in P_{k}$ and $S(\rho||\sigma) = -S(\rho)-\tr(\rho\log \sigma)$ is the quantum relative entropy, with $S(\rho)=-\tr(\rho\log \rho)$ the von Neumann entropy.
	The closest product state $\sigma_N$ of $\rho_N$ is the product of the reduced states of $\rho_N$ \cite{girolami2017, modi2010, bennett2011, szalay2015}, then 
	\begin{align}
	\label{eq:dis}
	I^{k\rightarrow N}(\rho_{N})&= S(\rho_{N}||\otimes^{m}_{i=1}\rho_{k_{i}}) \nonumber \\
	& =\sum_{i=1}^{m}S(\rho_{k_{i}})-S(\rho_{N}). 
	\end{align}
	As the physical system is invariant by spin  permutation, the evaluation of Eq.~\eqref{eq:dis} becomes simpler \cite{girolami2017} 
	\begin{equation}
	\label{eq:Dis}
	\begin{split}
	I^{k \rightarrow N}(\rho_{N})=& \left \lfloor N/k \right \rfloor S(\rho_{k})+\\
	&(1-\delta_{N\mod k,0})S(\rho_{N\mod k}) - S(\rho_{N}),
	\end{split}
	\end{equation}
	so that $\left \lfloor N/k \right \rfloor $ is the floor function and  $\rho_{N\mod k}$ describes the subsystem with $N\mod k$ spins. 
    %\sout{ In the particular case of $k=1$,}
	 %\begin{equation} \label{eq:totalcorr}
	%I^{1 \rightarrow N}(\rho_{N})=N S(\rho_{1}) - S(\rho_{N})
	%\end{equation}
	%\sout{ describes the total correlations presented in the system, \textit{i.e.,} how close is the state $\rho_N$ from the totally uncorrelated state. }
	%Therefore, we arrive at the following simpler form for the correlations.

	The GMC's of order $k$, denoted by $I^k$, are genuine correlations among $k$ subsystems of the whole system and can be calculated as the difference between the correlations of order higher than $k-1\rightarrow N$ and those of order higher than $k\rightarrow N$, 
	\begin{equation}
	\label{eq:Corr}
	I^{k}(\rho_{N})=I^{k-1 \rightarrow N}(\rho_{N})-I^{k \rightarrow N}(\rho_{N}).
	\end{equation}

	\textbf{Genuine Multipartite Correlations - } Summarizing, the GMC's in permutationaly invariant systems are given by
	\begin{eqnarray}
	\label{eq:gmc}
	I^k ( \hat \rho_N) & = &  \left \lfloor N/(k-1) \right \rfloor S(\rho_{k-1}) -
	\left \lfloor N/k \right \rfloor S(\rho_{k}) \nonumber   \\
	& &+ (1-\delta_{N\mod k-1,0})S(\rho_{N \mod k-1}) \nonumber \\ 
	& & - (1-\delta_{N\mod k,0})S(\rho_{N \mod k}) \nonumber \\
	& & 
	\end{eqnarray}
for $k > 1$. For $k=1$ we have instead,
\begin{equation} \label{eq:totalcorr}
	I^{1}(\rho_{N}) \equiv N S(\rho_{1}) - S(\rho_{N}),
	\end{equation}
	which describes the total correlations presented in the system, \textit{i.e.,} how close is the state $\rho_N$ from the totally uncorrelated state. 
	
\textbf{Quantum Fisher Information - } The GMC's encompass both classical an quantum multipartite correlations among the subsystems. It would be interesting to discriminate both types of correlations in the phases of the model. Therefore we also study the QFI, which is a well known witness of $k$-partite quantum entanglement \cite{pezze2009,hyllus2012,toth2012}. Specifically, we compute the maximum QFI optimized over the global spin observables $\hat S_\alpha$. The optimum (maximum) QFI for a general mixed state is given by the maximum eigenvalue of the $3\times3$ matrix:
	\begin{equation}
	\label{eq:maxQFI}
	[ \Gamma ]_{kl}=2\sum_{i,j}\frac{(p_i-p_j)^2}{p_i+p_j}\bra{j}\hat{S}_k/2\ket{i}\bra{i}\hat{S}_l/2\ket{j},   
	\end{equation}
	with $k,l=x, y, z$, $p_i+p_j>0$, and $\hat \rho=\sum_i p_i\ket{i}\bra{i}$ is the spectral decomposition of the state. We denote the maximum of the QFI as $F_{\max}$. The entanglement witness feature comes from a simple inequality: the state $\hat \rho$ has k-partite entanglement if $F_{\max}(\hat \rho)/N>(k-1)$. Otherwise, if $F_{\max}<N$ it is not possible to conclude that the system is not entangled, since the witness $F_{\max}$ may have just failed to capture it, behaving as a flawed entanglement witness for such a system.

%%%%%%%%%%%%%%%%%%%%%%%%%%%%%%%%%%%%%%%%%%%%%%%
	\section{Non-equilibrium Steady States}
	%%%%%%%%%%%%%%%%%%%%%%%%%%%%%%%%%%%%%%%%%%%%%%%
	\label{sec.ness}
	
	\begin{figure*}
    \includegraphics[width=.48\textwidth]{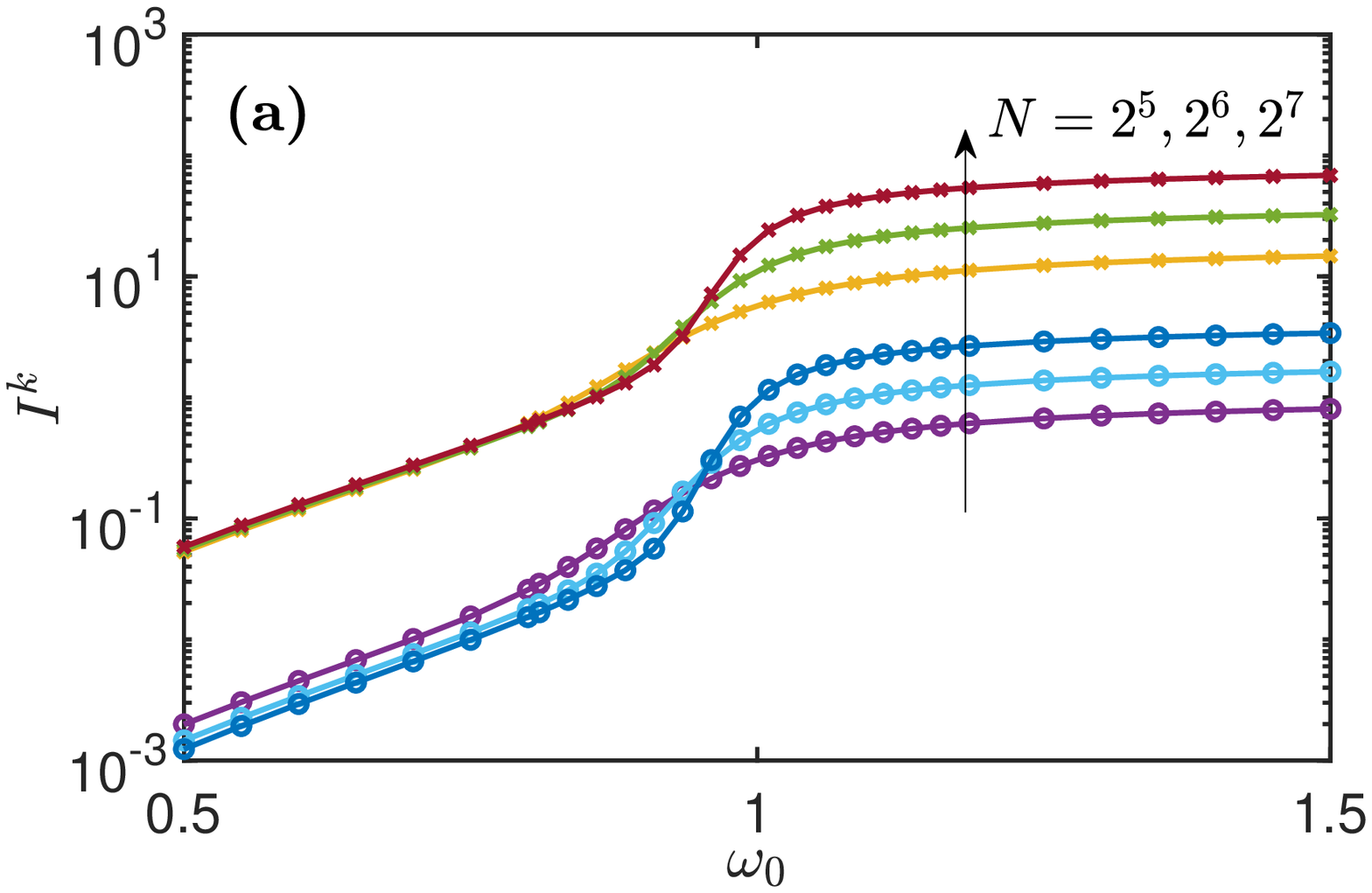}
    \includegraphics[width=.47\textwidth]{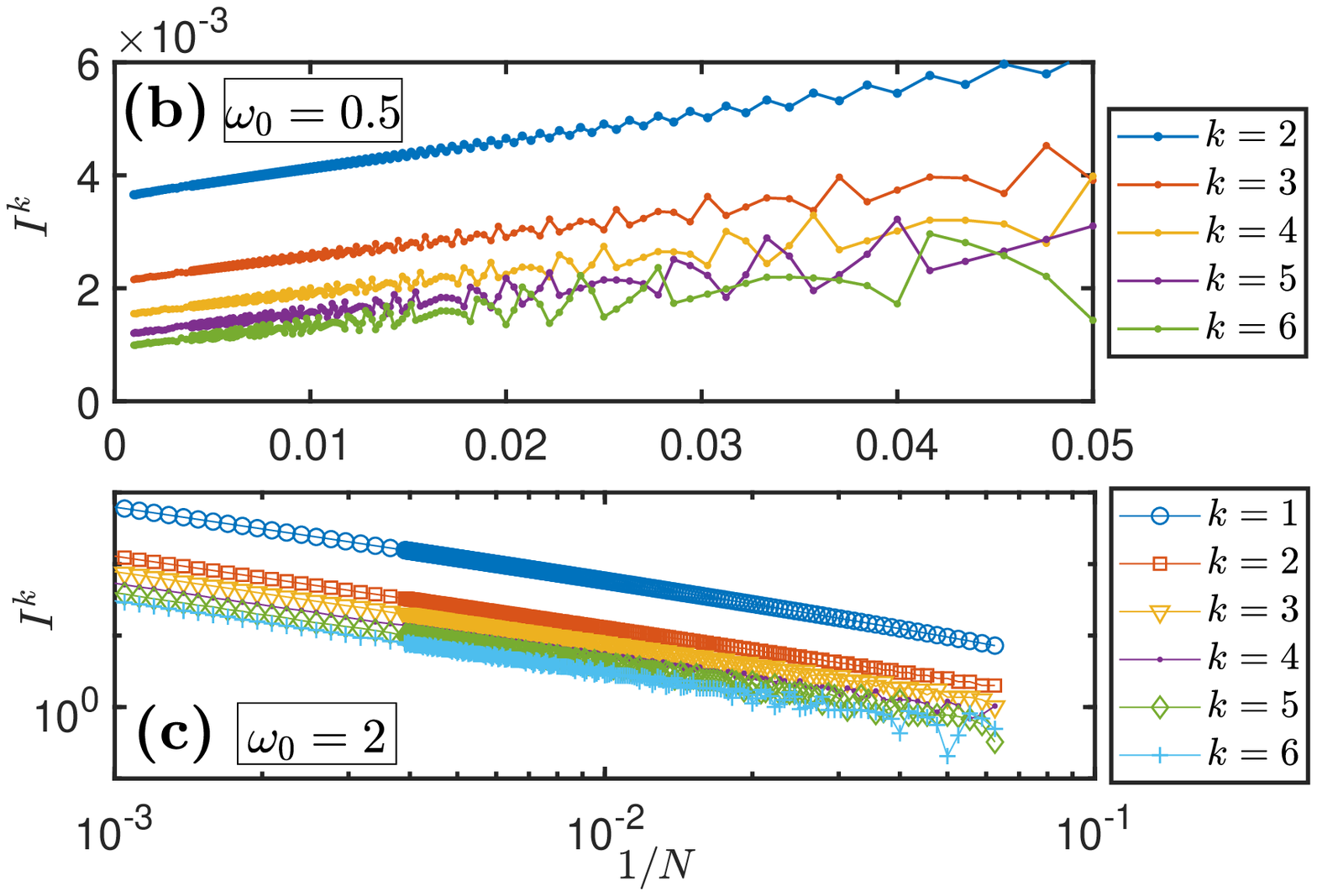}
    \includegraphics[width=.48\textwidth]{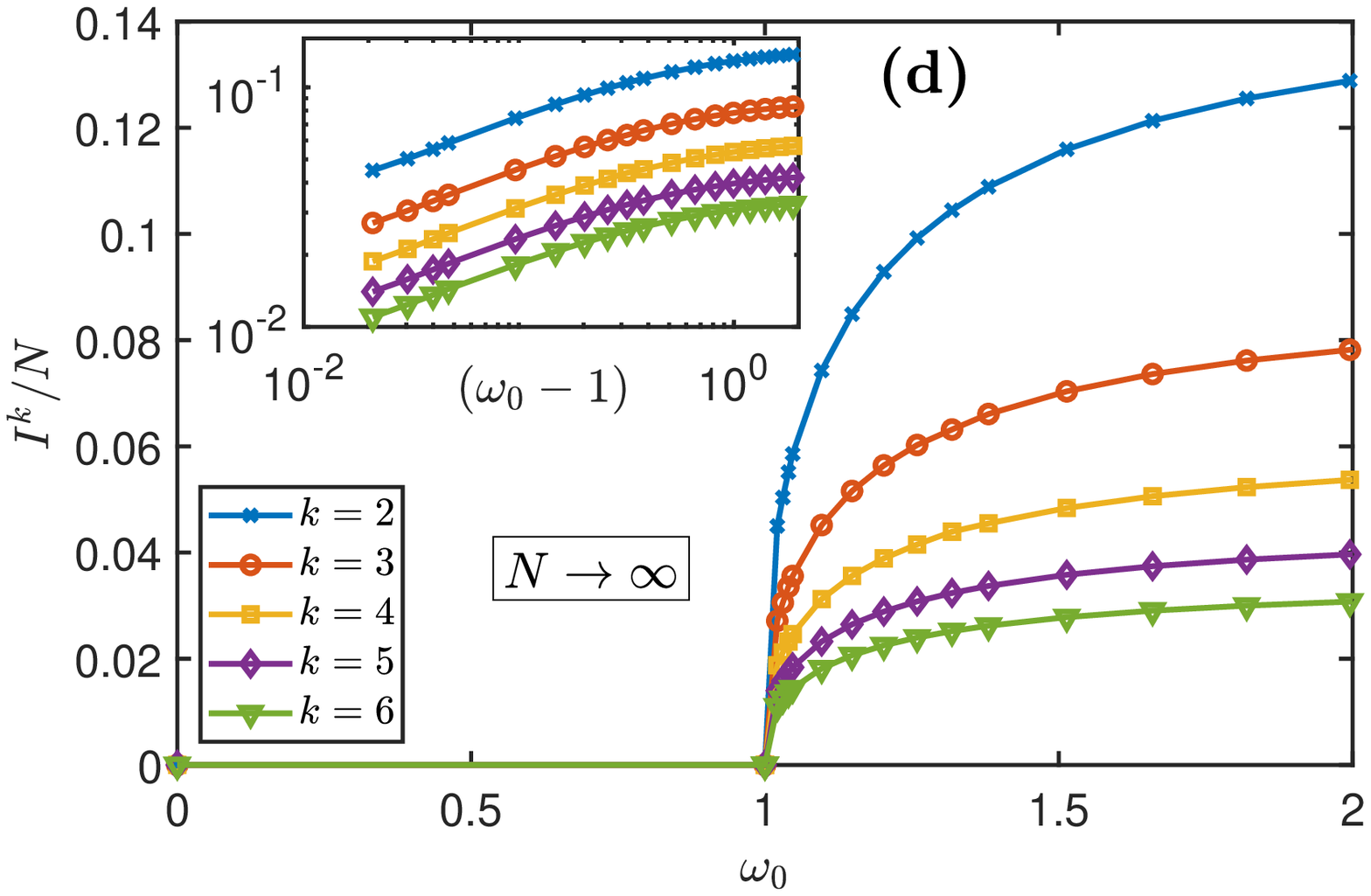}
    \includegraphics[width=.48\textwidth]{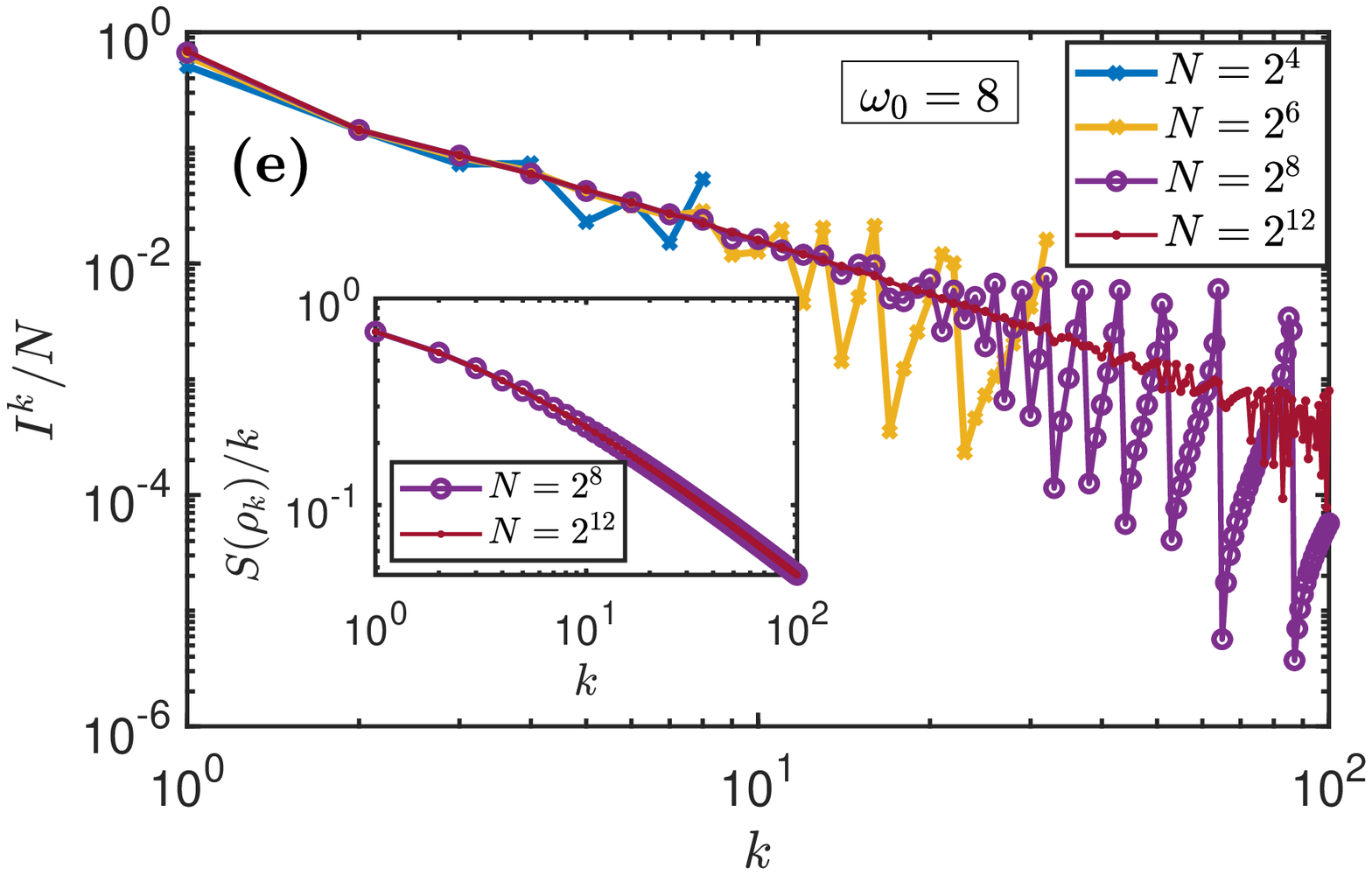}
		\caption{Genuine multipartite correlations (GMC's) for the NESS of the system, considering  different $k$-partitions and along the full phase diagram of the system. \textbf{(a)} We show the GMC's for $k=1$ (crosses) and $k=6$ (circles) and different system sizes, showing their behavior in the two different phases of the model: while in the ferromagnetic phase it has subextensive, but finite, correlations (i.e., 
		$\lim_{N \rightarrow \infty} I^k/N = 0$ with 
		$\lim_{N \rightarrow \infty} I^k \neq 0$ - see panel \textbf{(b)} for the finite size scaling), in the BTC phase it features extensive multipartite correlations (panel \textbf{(c)}). We show in \textbf{(d)} the correlations in the thermodynamic limit within our truncated ansatz approach, corroborating this behavior for all $k$ partitions. In the inset of the figure we show the quantum phase transition at $(\omega_0)_{\rm c} = 1$, for which the multipartite correlations of all orders display a power-law critical behavior, with $I^k/N \sim (\omega_0 - (\omega_0)_{\rm c})^\beta$ and exponent $\beta  \sim 0.3$. In panel \textbf{(e)} we show structure of GMC's with the $k$ orders, in the BTC phase. We see a power-law decay of genuine correlations with $k$ (the inset display the raw reduced entropy $S(\rho_k)/k$).
		} 
		\label{fig:NESSvshx}
    \end{figure*}

	We first study the structure of the correlations in the non-equilibrium steady states (NESS) of the system. The NESS is reached after long times 
	($t \rightarrow \infty$) of the evolution dictated by master equation Eq.\eqref{eq:evo}, or in other words, it corresponds to the solution of the Lindbladian differential equation $\frac{d}{dt}\hat{\rho}_{\rm NESS}=\mathcal{\hat{L}}[\hat{\rho}_{\rm NESS}]=0$. The exact expression for the NESS of this model is known~\cite{hannukainen2018}, and given by (see Appendix \ref{appendix})
	\begin{equation}
     \hat \rho_{\rm NESS} = \frac{1}{\mathcal{N}} \hat \eta \hat \eta^\dagger, 
     \end{equation}
    with  
    \begin{equation} \label{eq.eta.ness}
    \hat \eta = \sum_{j=0}^{N} (\hat S_-/g^*)^j.
    \end{equation}
    where $g=i\omega_0N/2$ and the normalization constant is 
    \begin{align}
        \mathcal{N}=N\sum_{j=0}^{N}\frac{1}{|\tilde{g}|^{2j}}\sum_{m=0}^{j}\frac{(-1)^m}{2m+1}\binom{j}{m},
    \end{align}
    for which $\tilde{g}=2g/N$.

\subsection{Genuine Multipartite Correlations}

 Despite the exact expressions for NESS, it is not immediate how to extract (compute) the genuine correlations from it. For moderate finite system sizes (of the order of $N \lesssim 2^{12}$) one can write explicitly the state numerically and directly compute its properties. For the thermodynamic limit ($N \rightarrow \infty$), however, we perform a truncation over the $\hat \eta$ operators and compute analytically its $n$-body correlations up to $n \sim 10$,  therefore also its GMC's of order $n$ (see Appendix \ref{appendix} for details).
 The truncation approach is based on considering only a few ``$\ell_{\tr}$'' terms in the sum of excitations in  Eq. \eqref{eq.eta.ness}. We have observed that the correlations within this truncated ansatz converge exponentially fast to the exact results for increasing $\ell_{\rm tr}$, thus leading to negligible inaccuracies even for finite $\ell_{\rm tr}$ and to a reliable computation of the GMC's in the thermodynamic limit.
 
 We show our results for the GMC's in Fig. \ref{fig:NESSvshx}. We first observe that the NESS has nonzero GMC's along the two different phases of the model. 
 In the ferromagnetic phase ($0 \leq \omega_0 <1$), however, the spins have much weaker correlations among each other compared to the BTC phase. While the GMC's in the ferromagnetic phase are subextensive with system size (i.e., $\lim_{N\rightarrow \infty} I^k/N = 0$, but nevertheless nonzero $\lim_{N\rightarrow \infty} I^k \neq 0$), within the BTC phase the spins feature extensive GMC's. See Fig. \ref{fig:NESSvshx}b and Fig. \ref{fig:NESSvshx}c for the finite size scaling of the correlations along the two phases, and Fig. \ref{fig:NESSvshx}d for the correlations in thermodynamic limit, obtained through the truncation ansatz approach. As usual in spontaneous symmetry breaking theory, the different ordered phases arise due to the interacting nature of their constituents, correlations play therefore a fundamental role in their characterization. We see that this behavior is  corroborated in the BTC of our model, leading to the extensivity of the GMC's. It is interesting to put these results in context to different forms of TC's, such as those occurring in closed systems. In this case one can stabilize discrete time crystals (DTC's), for which correlations are also expected. A non-null mutual information between distant spins (in the Floquet eigenstates) is shown along the DTC's and tends to zero as one approaches the invariant (non-TC) phase \cite{yao2017,else2016}. We recall that the mutual information between two spins quantify the total correlations between them, i.e., quantum and classical correlations \cite{henderson2001classical}, in the same spirit as GMC's.
   We show in Fig.\ref{fig:NESSvshx}d-(inset panel)  that the GMC's display a second order phase transition at the critical point $(\omega_0)_c=1.0$, with a power-law singularity $I^k/N \sim (\omega_0 - (\omega_0)_{\rm c})^\beta$ and exponent $\beta  \sim 0.3$. %\textcolor{red}{The subextensivity in the ferromagnetic phase and extensivity in BTC poses the question if this is a characteristic of continuous phase transitions, as seems to happen also in the quantum phase transition of the Lipkin-Meshkov-Glick model \cite{lourenco2020}. This question is worth of investigation, but no progress in this sense will be made here.} 
  %\purple{  All orders of GMC's have a similar qualitative behavior along the phase diagram of the model in the thermodynamics limit. This fact can be related to the collective nature of their couplings (Eq.\eqref{eq:evoequation}), which preserves particle permutation symmetry and therefore different partitions can share much similar properties.  }

    \begin{figure}
    \includegraphics[width=.49\textwidth]{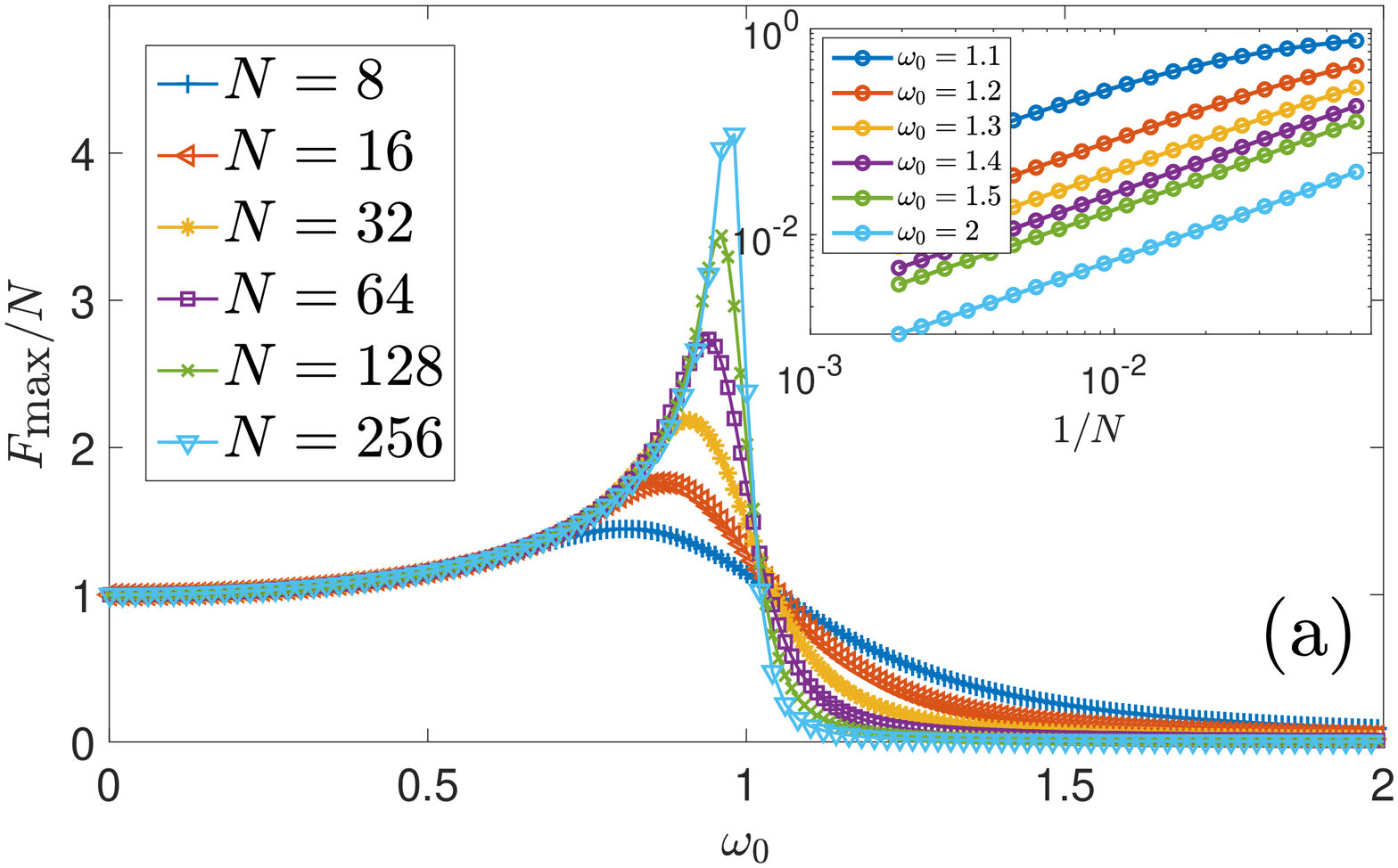}
    \includegraphics[width=.49\textwidth]{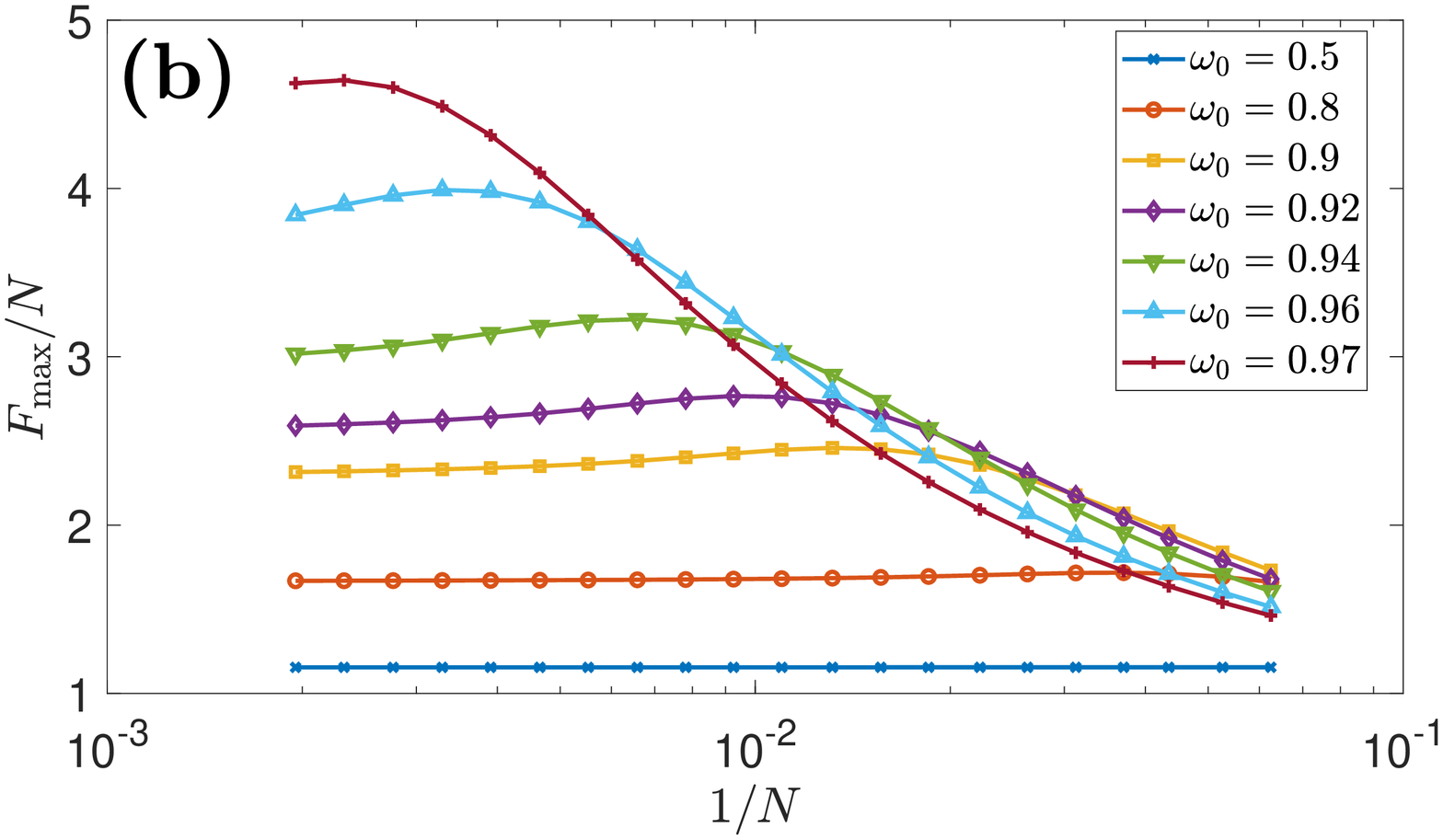}
	\caption{Maximum value of the quantum Fisher information ($F_{\textnormal{max}}$) for the NESS of the system along the phases of the model. We show in panel \textbf{(a)}  $F_{\max}$ for varying couplings and different system sizes. The QFI witnesses multipartite entanglement only in the ferromagnetic phases, reaching its peak around the critical coupling. Panel \textbf{(b)} shows the finite-size scaling of the QFI in the ferromagnetic phase. The BTC phase (inset of panel \textbf{(a)}) presents a vanishing power-law decay with system size.  
	}
	\label{fig:FmaxNESSvshx}
	\end{figure} 

   The structure of GMC's in the NESS shows a peculiar behavior, see Fig. \ref{fig:NESSvshx}e. 
   The hierarchy among $k$-partite orders $I^k>I^{k+1}$ is almost preserved, except for partitions with $k$ not being a multiple of $N$. In this case the multipartite correlations display an atypical oscillatory behavior - mathematically, this could also be attributed to the role of the floor function in Eq. \eqref{eq:Dis}, which brings an additional contribution. Nonetheless, in the thermodynamic limit they are absent, and we see a power-law hierarchy $I^k \sim k^{-\alpha}$, with $\alpha >0$. Moreover, from the simpler reduced density matrix entropies (inset panel) - which do not necessarily corresponds to multipartite correlations - there is no oscillatory behavior. It is interesting try reasoning these oscillations on top of monogamy of correlations between the subsystems; since they are permutationally invariant, breaking the symmetry might induce atypical behavior. This subject must be deeper understood, but stands as an interesting perspective. We also recall that such a behavior has also been found in other physical systems \cite{lourenco2020,calegari2020}. 
   %\textcolor{red}{The fact the all orders of GMC's have the same pattern in each phase of matter comes from the similar way in which each spin interacts with the driving field, as well as its interaction with a common reservoir. This preserves the particle permutation symmetry and synchronizes the behavior between them. In this way, each partition of the system is behaving in a similar manner, as captured by GMC's or QFI, despite of the intensity of the correlation diminishes in average with the increase of the partition.}
    \begin{figure}
    	\includegraphics[width=.49\textwidth]{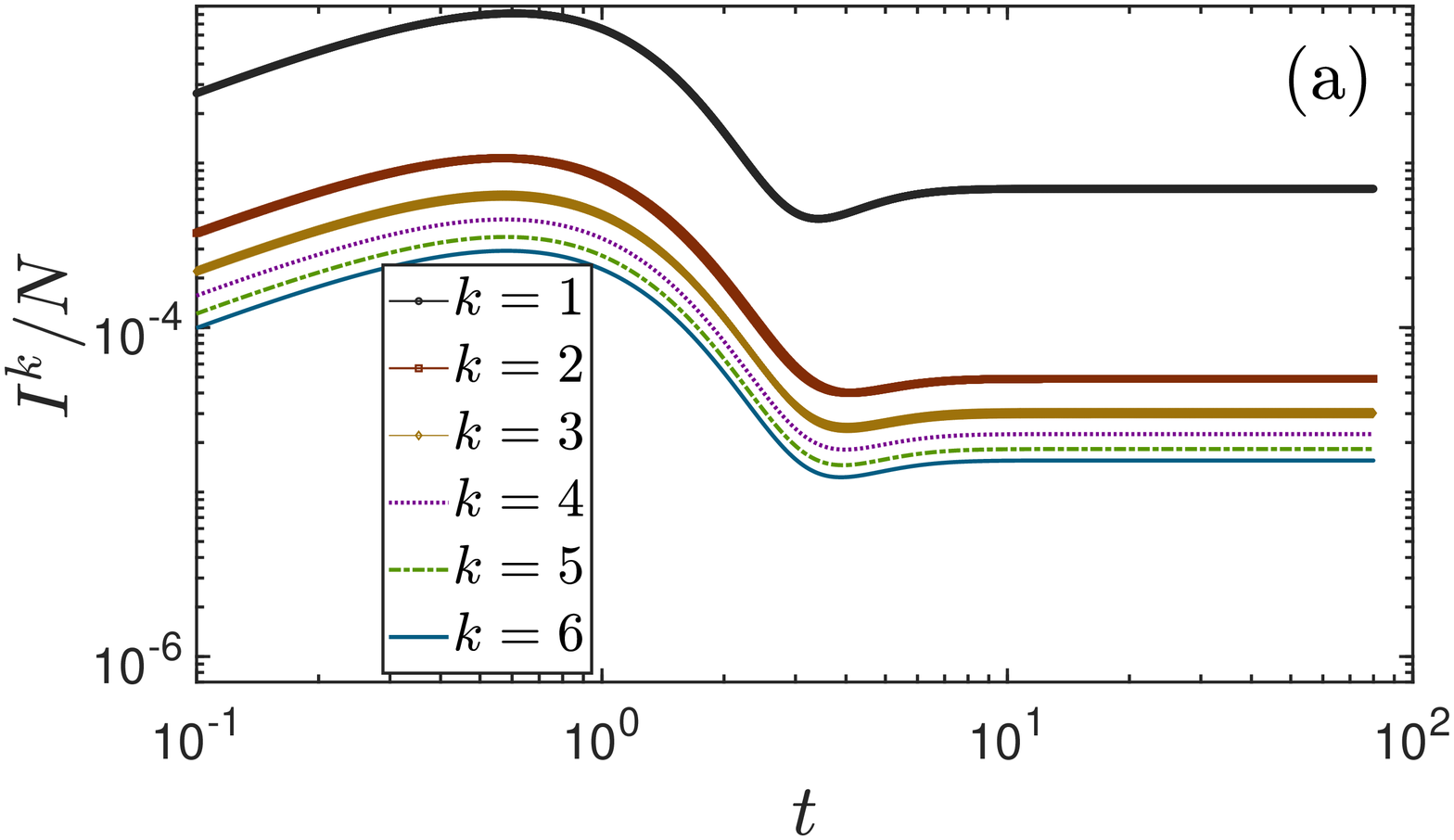}
    	\includegraphics[width=.49\textwidth]{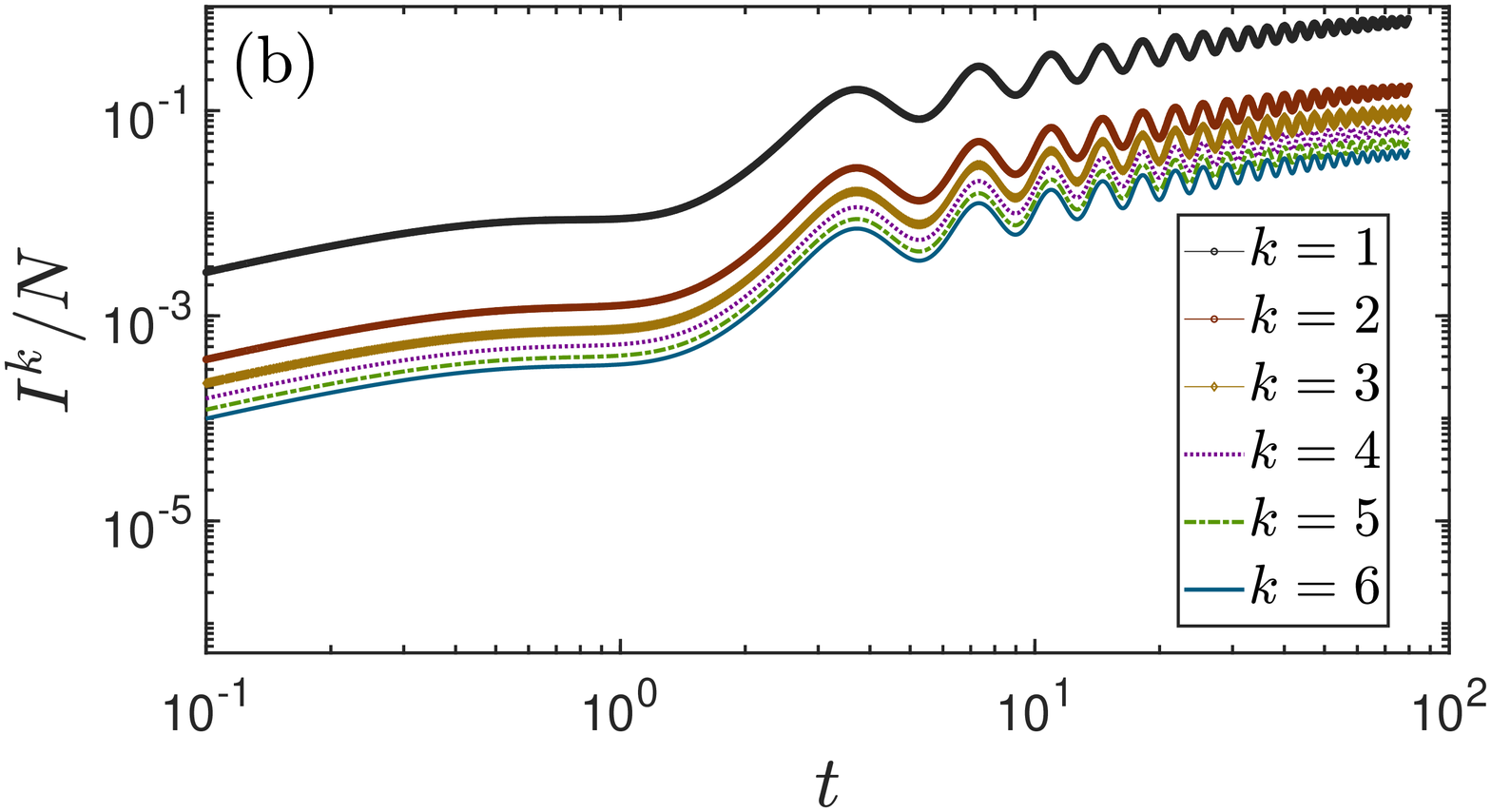}
    	\caption{Dynamics of GMC's for different $k$'th orders and along the two different phases of the model. In panel \textbf{(a)} we show the dynamics in the ferromagnetic phase with $\omega_0=0.5$, and   panel \textbf{(b)} shows the dynamics in the BTC  phase with $\omega_0=2$. The number of $1/2-$spins in the system is $N=120$.} \label{fig:STL120w05}
  \end{figure} 
   
 %%%%%%%%%%%%%%%%%%%%%%%%%%%%%%%%%%%%%%%%  	
 \subsection{Quantum Fisher Information}
 %%%%%%%%%%%%%%%%%%%%%%%%%%%%%%%%%%%%%%%%

 In an attempt to discriminate the classical and quantum roots of the multipartite correlations, we examine the QFI along the two phases of the model. We find, surprisingly, that the QFI witnesses quantum entanglement only in the ferromagnetic phase (the one with substantially weaker GMC's) - see Fig. \ref{fig:FmaxNESSvshx}a.
 For finite system sizes, the QFI in the ferromagnetic phase increases its value for increasing the coherent field $\omega_0$, witnessing $k$-partite entanglement till reaching its maximum value close to $(\omega_0)_c$. A similar result was also observed for two-spins entanglement captured by the negativity quantifier \cite{hannukainen2018}. The QFI, however,  also witnesses  entanglement between larger groups of spins - a subextensive number of spins. Precisely, it witness $k \sim \mathcal{O}(1)$-partite entanglement as shown by a finite-size scaling  analysis (see Fig. \ref{fig:FmaxNESSvshx}b). Moreover, when close to the quantum phase transition, the maximum witnessing entanglement occurs for a finite system size $N^*$, which depends on the couplings parameters.

     On the other side, for the BTC phase there is not witnessing of entanglement and the QFI decays as a power-law with the system size (inset panel of Fig. \ref{fig:FmaxNESSvshx}a).
     Nevertheless, GMC's are known extensive in this phase,  so these results point towards either (i) to the classicality of these correlations, or (ii) simply a failure of QFI to witness entanglement, or (iii) a third route with possibly discord-like quantumness of correlations (nor classical neither entanglement).
      Ultimately we are not able to fully discriminate these three possibilities, it is nevertheless interesting to discuss them in connection to the purity and coherence of the NESS. The NESS in the BTC phase are known as highly mixed states \cite{hannukainen2018,piccitto2021}, for which the computation of entanglement is in general a highly nontrivial task. This property may related to the failure of the QFI to witness its entanglement. On the other hand, besides highly mixed, the NESS have a peculiar structure in the Dicke basis described by an "almost diagonal" density matrix (not fully diagonal though), i.e., with a very low coherence in the Dicke basis (see Appendix.\eqref{app:coherence}, for a detailed discussion).  In the extremal limit with $\omega_0/\kappa \rightarrow \infty$ the coherence in the Dicke basis tends to zero and the density matrix becomes diagonal in such a basis. States in this diagonal form were studied in Refs.\cite{wolfe2014,yu2016separability,tura2018separability} and shown not having entanglement, however, they do have quantum-discord correlations \cite{santos2016}. If the properties of this extremal limit can be extended through all BTC phase, we would conclude that the phase is non-entangled, but quantum-discord correlated. We cannot assure this fact, however, since an extremal limit may be a  singular point and not fully descriptive for the whole phase properties.

    %%%%%%%%%%%%%%%%%%%%%%%%%%%%%d
	\section{Dynamics}
	%%%%%%%%%%%%%%%%%%%%%%%%%%%%%
	
	\label{sec:res}
	
	In this section we discuss the dynamical properties of the correlations through GMC's quantifier and  QFI witnesses, along the different phases of the model.
	
	\subsection{Genuine Multipartite Correlations}

	The behavior of the GMC's during the dynamics is shown in Fig. \ref{fig:STL120w05}. For short times, $t \ll 1$, the GMC's grow according to a power-law for both phases of the model and any $k$'th order.  In the ferromagnetic phase, after this initial transient time the GMC's quickly decay to a constant value in which the system goes towards its NESS - Fig. \ref{fig:STL120w05}a. The behavior is similar for all orders of $k$, differing only in the amount of correlation, i.e., decreasing its value for higher values of $k$. 
	\begin{figure}
        \includegraphics[width=.48\textwidth]{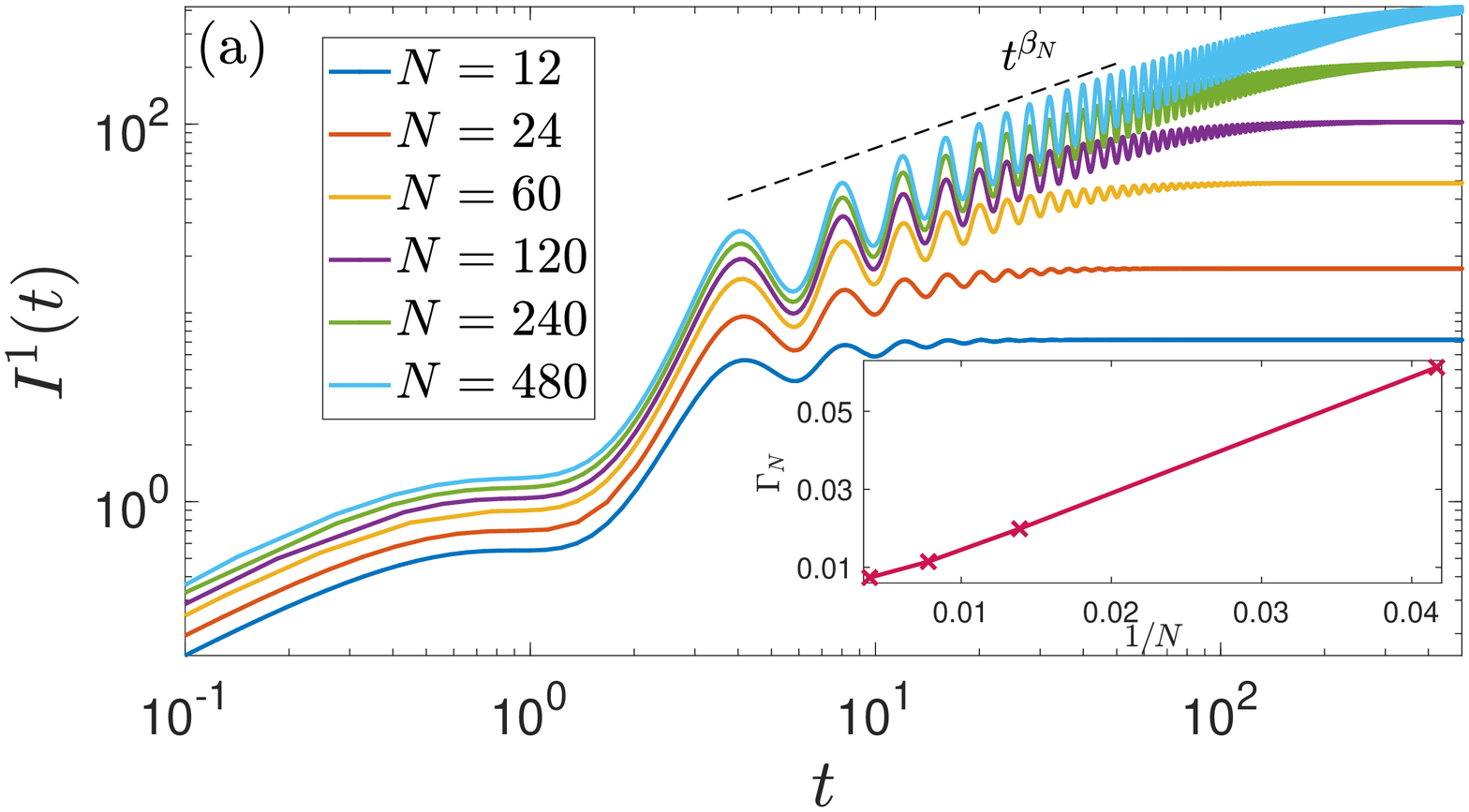}
        \includegraphics[width=.49\textwidth]{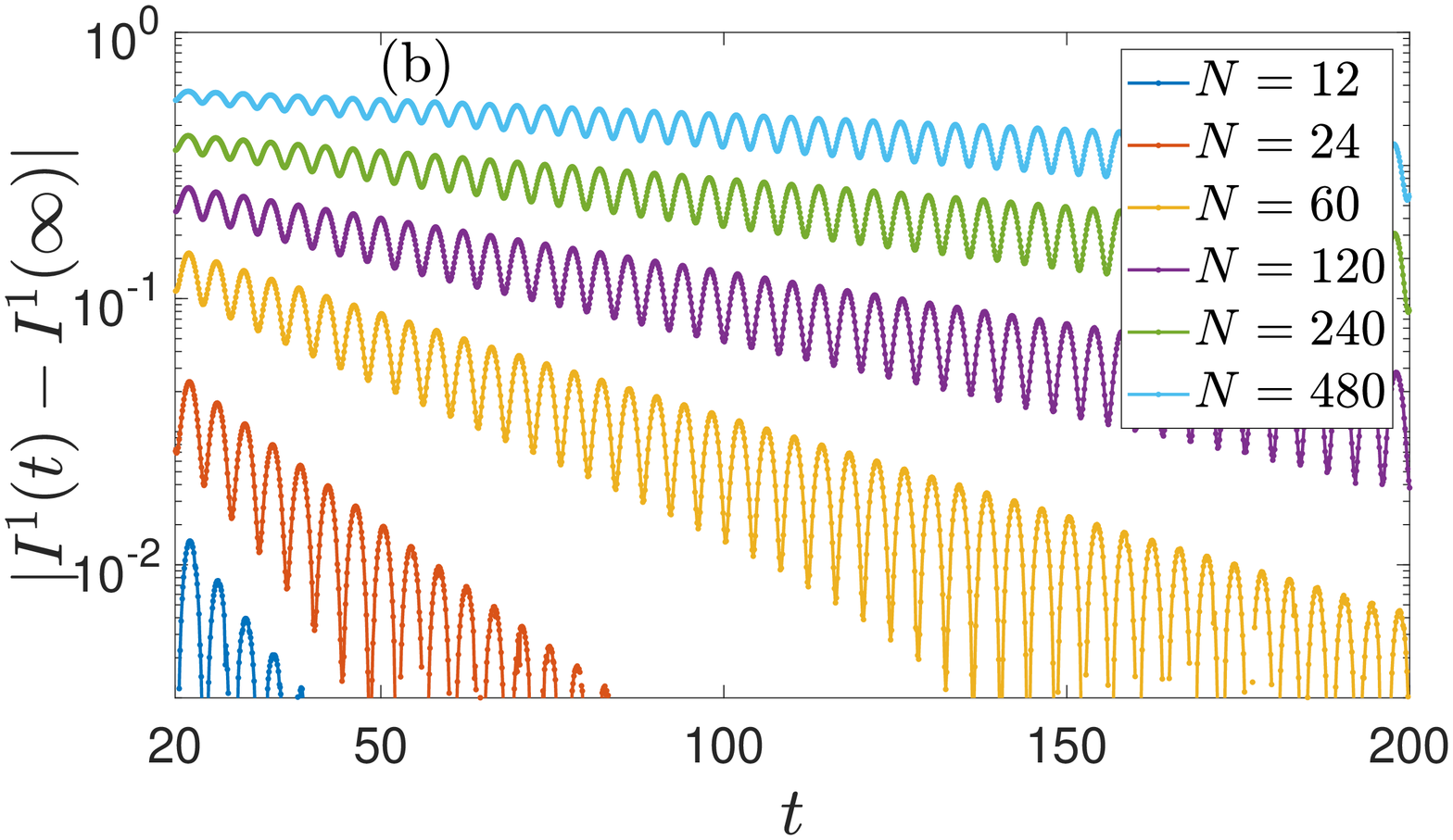}
        \caption{We show in panel \textbf{(a)} the dynamics of the total correlations $I^1(t)$ within the BTC phase, for $\omega_0=2$ and different system sizes. After an initial transient time, the correlations osillate and grow according to a power-law, which are however, damped due to finite size effects according to Eq.\eqref{eq:Ik.dynamics}. In panel (b) we show the approach of the correlations to their steady state values, in a log-linear scale, highlighting the exponential behavior with a decreasing damping rate for increasing system sizes. The damping rate for different system sizes is shown in the inset of panel (a), displaying a divergence of the correlations lifetime in the thermodynamic limit. 
        } \label{fig:STmultL}
    \end{figure} 
	
    On the other hand, in the time crystal phase,  the spins after the initial transient time (i.e., for $t\gtrsim 1$) turn into a periodic motion leading to an oscillatory and growing dynamics for the correlations - see Fig. \ref{fig:STL120w05}b. The correlation growth (and oscillations) are damped due to the interaction with the environment, with a lifetime depending both on the system size as on the system couplings.
	
    We show in Fig. \ref{fig:STmultL}a the dependence of their dynamics with the system size. Specifically, we find that the correlations behave as
    \begin{equation} \label{eq:Ik.dynamics}
     I^k(t) \sim t^{\beta_N} e^{-\Gamma_N t} \cos(\nu t) + I^k_{\rm NESS},
    \end{equation}
    where $I^k_{\rm NESS} \equiv I^k(t \rightarrow \infty)$ is its non-equilibrium steady state value, $\beta_N$ characterizes the power-law growth, $\nu$ is the frequency of the oscillations, and $\Gamma_N^{-1}$ corresponds to the lifetime of the dynamics.
    \begin{figure}
	\includegraphics[width=.49\textwidth]{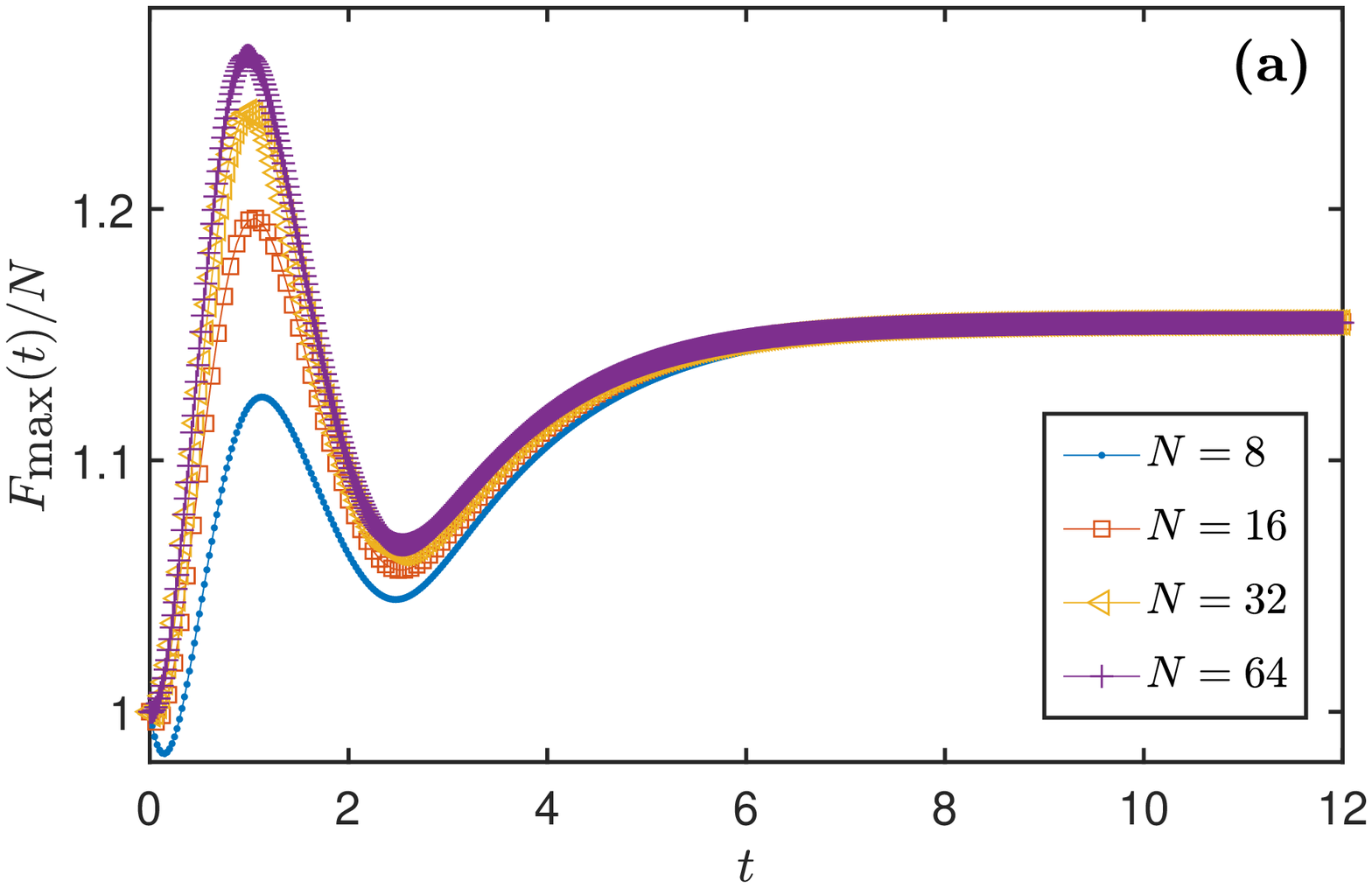}
	\includegraphics[width=.5\textwidth]{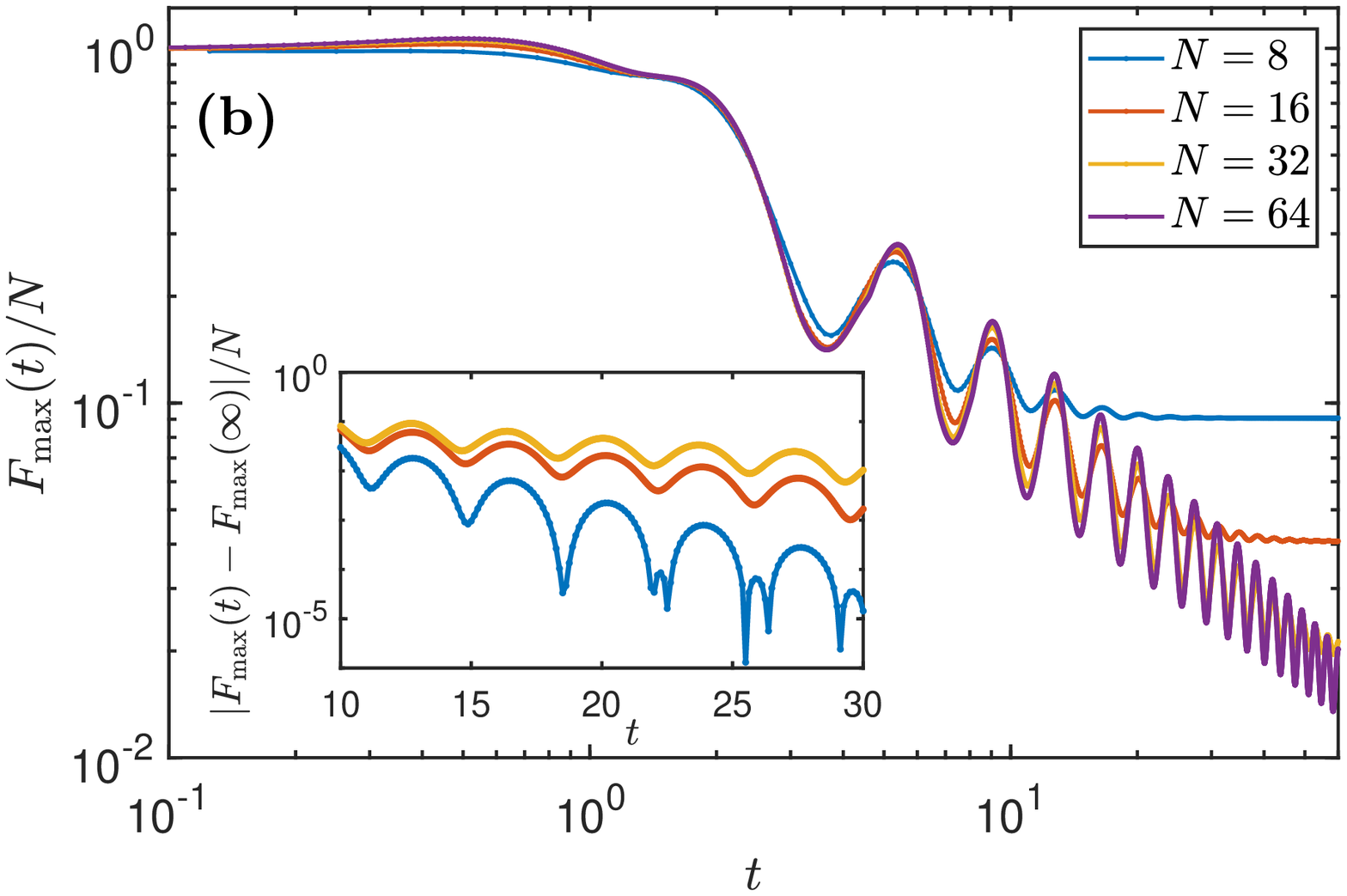}
    \caption{ The maximum value of QFI - $F_{\rm {max}}$ - during the dynamics for the two phases of the model and different system sizes. In panel \textbf{(a)} we show the dynamics in the ferromagnetic phase ($\omega_0=0.5$), for which one can observe that genuine multipartite entanglement between two spins are generated and maintained, even for long times. In the panel \textbf{(b)}, for the BTC phase  with $\omega_0=2.0$, we see that QFI witnesses entanglement only for short times, and show a rather oscillatory decay towards its steady state value. The lifetime of the this oscillatory dynamics diverge in the thermodynamic limit, as can be seen from the log-linear inset panel.}
    \label{fig:Fmaxvst}
	\end{figure}
   
    While for intermediate times ($t\gtrsim 1$) it is clear the power-law growth (dominant term in the above equation), for long times the exponential term plays the major role  damping the correlations.
    We can extract the lifetime of the dynamics from its long time behavior, $I^k(t)-I^k_{\rm NESS} \sim e^{-\Gamma_N t}$ for $t\gg 1$, whose exponential depends on the decay rate $\Gamma_N$ shown in Fig. \ref{fig:STmultL}b (or from a nonlinear fit of the dynamics). We obtain that the lifetime of the correlations diverges algebraically with the system size ($\Gamma_N \sim N^{-1}$, see inset panel of Fig. \ref{fig:STmultL}a, highlighting the intermittent growth of GMC's in the thermodynamic limit.

    %%%%%%%%%%%%%%%%%%%%%%%%%%%%%%%%%%%%%%%%%%%%%%%%
	\subsection{Quantum Fisher Information}
	%%%%%%%%%%%%%%%%%%%%%%%%%%%%%%%%%%%%%%%%%%%%%%%%

	We show in Fig. \ref{fig:Fmaxvst} the dynamics of the maximum QFI $F_{max}$ along the two phases of the model. We see some similarities as compared to GMC's: while the QFI quickly saturates to a constant value in the ferromagnetic phase, it displays persistent dynamics (with a size dependent damping rate) within the BTC phase. It is surprisingly to notice that the QFI witnesses entanglement in its dynamics only in the ferromagnetic phase. In the BTC phase, apart from short times $t \lesssim 1$, we have that  $F_{\rm max}<N$ does not truly witness any 
	multipartite entanglement correlations. We recall however that QFI is a witness, and not a full quantifier of entanglement, thus we can not discard the possible presence of such correlations along the dynamics of the phase. %\sout{\textcolor{red}{Although, we know that for large $N$ and ...}}
	%approaching the NESS the state of the system tends to a diagonal in the Dicke basis and indicating that is not entangled}}
	%\brown{FI: essa discussão foi adicionada na seção de NESS, desta forma não tem necessidade repetí-la aqui]}. 
	Nevertheless, the QFI in the BTC captures the oscillatory and persistent (in the thermodynamic limit) behavior. We see in Fig. \ref{fig:Fmaxvst}b that the QFI decays towards its steady state value according to
	 \begin{equation}\label{eq:QFI.dynamics}
	  F_{\rm max}(t)/N \sim t^{-\alpha_N} e^{-\overline{\Gamma}_N t} \cos(\nu t) + F_{\rm max,NESS},
	 \end{equation}
	 with $F_{\rm max,NESS}  \equiv F_{\rm max}(t \rightarrow \infty)$ being its non-equilibrium steady state value, $\alpha_N$ characterizes the power-law decay, $\overline{\Gamma}_N$ is the effective decay rate, and $\nu$, as for GMC's, is the BTC frequency.  As for GMC, the effective decay rate vanishes algebraically with the system size $\overline{\Gamma}_N \simeq N^{-1}$.  The inset of Fig. \ref{fig:Fmaxvst}b highlights the divergence of the lifetime for the oscillatory dynamics in the thermodynamic limit.

   \section{Conclusions and Perspectives}\label{sec:conclusion}
   %%%%%%%%%%%%%%%%%%%%%%%%%%%%%%%%%%%%%%%%%%%%%%%%%%%%%%%%%%%%%%
    
    In summary, we studied GMC and QFI in a many-body system composed by spin $1/2$ particles interacting with an environment that can be in a ferromagnetic or BTC phase, depending on the coupling parameters. We found for the NESS of the system that all orders of the GMC are extensive with system size in the BTC phase, while are subextensive in the ferromagnetic phase. 
    Furthermore, the GMC show a second order phase transition between these phases with associated critical exponent $\beta \approx 0.3$. Given that such quantifier accounts for classical as well as quantum correlations, we also analysed the QFI in order to witness multipartite entanglement between the spins of the system. Surprisingly, the QFI  vanishes with the system size in the BTC phase, detecting entanglement only in the ferromagnetic phase. We are therefore not fully able to resolve the nature of the correlations present at the NESS of the BTC phase, although we have indications that it has at least quantum discord correlations - from the analysis of the extreme case $\omega_0/\kappa \rightarrow \infty$. The dynamics of both GMC and QFI show in the thermodynamic limit, and only in such a limit, a persistent oscillatory dynamics. While the GMC display the persistent oscillatory behavior around a mean algebraic growth Eq. \eqref{eq:Ik.dynamics}, for all  $k$ orders of genuine correlations, the persistent oscillations in the QFI appears around a mean algebraic decay Eq. \eqref{eq:QFI.dynamics}.
    
    An interesting perspective stands for a deeper investigation of the nature of the correlations (classical or quantum) and their role in the BTC phase. Furthermore, the time-crystal is nevertheless shown strongly correlated, thus enabling the possibility of exploiting such correlations to improve thermal machines by reversing the heat flow \cite{micadei2019}. Also, as the structure of GMC's presents a peculiar behavior dependent of $k$ in the NESS, also noticed in different systems \cite{lourenco2020,calegari2020}, it would be worth an investigation of their roots in connection to these different models  where the same behavior was observed, from the role of floor function in the quantifier to monogamy/frustration of their correlations. Lastly, the subextensivity of the GMC's in the ferromagnetic phase and its extensivity in the BTC phase poses the question if this is a characteristic of continuous phase transitions, as it seems to happen also in the second-order quantum phase transition of the Lipkin-Meshkov-Glick model \cite{lourenco2020}. This question is worth of investigation and we leave it as an a future work.

	\begin{acknowledgments}
		The authors acknowledge financial support from
		the Brazilian funding agencies Coordenação
		de Aperfeiçoamento de Pessoal de Nível Superior
		(CAPES), Conselho Nacional de Desenvolvimento
		Científico e Tecnológico (CNPq), Fundação de Amparo à Pesquisa e Inovação do Estado de Santa Catarina (FAPESC), and Instituto
		Nacional de Ciência e Tecnologia de Informação
		Quântica (CNPq INCT-IQ (465469/2014-0)).
		F.I. acknowledges the financial support of the Brazilian funding agencies CNPq (Grant No. 308205/2019-7) and FAPERJ (Grant No. E-26/211.318/2019).
	\end{acknowledgments}
	
	%%%%%%%%%%%%%%%
	\begin{appendix}
	%\appendix
	%%%%%%%%%%%%%%%
	\begin{widetext}
	\section{Analytic calculation of the NESS}
	\label{appendix}

    \subsection{Steady state of Dicke model}
    The steady state of Dicke model can be written as \cite{hannukainen2018}
    \begin{equation}
        \label{eq:ssdm}
        \hat{\rho}_{ss}=\frac{1}{D}\hat{\eta}\hat{\eta}^{\dagger},
    \end{equation}
    where
    \begin{equation}
        \label{eq:etass}
        \hat{\eta}=\sum_{j=0}^{N}\left(\frac{\hat{S}^{-}}{g^*}\right)^j,
    \end{equation}
    with $g=i\omega_0\frac{N}{2}$ and the normalization constant is 
    \begin{align}
        D=\sum_{j,l=0}^{N}\Tr{\left(\frac{\hat{S}^-}{g^*}\right)^j\left(\frac{\hat{S}^+}{g}\right)^l}=\sum_{j=0}^{N} \frac{\Tr{ (\hat{S}^-\hat{S}^+)^j }}{(g^*g)^j}.
    \end{align}
    In order to calculate the NESS in the thermodynamic limit it was made a truncation of the steady state in Eq. \eqref{eq:ssdm}. For that we used the following ansatz 
    \begin{eqnarray}
        \hat \eta &=& \sum_{j=0}^{\ell_{tr}} \left(\frac{\hat S^-}{g^*}\right)^j,
    \end{eqnarray}
    where $\ell_{tr}$ determines the order of the truncation ansatz.
    
    Since in the thermodynamic limit the multipartite correlations can be approximated by
    \begin{equation}
        \frac{I^k(\hat \rho)}{N} \sim \frac{S(\hat \rho_{k-1})}{k-1} - \frac{S(\hat \rho_{k})}{k},
    \end{equation}

    with ``$\sim$'' excluding finite-size effects, we can analyze the effect of truncation in $S_{\ell_{tr}}(\rho_k)/k$, which one was made with numerical computation. The results are shown in Fig. \ref{fig:GMC_var_ltruncation}. 
\begin{figure}[ht]
     \includegraphics[scale=0.3]{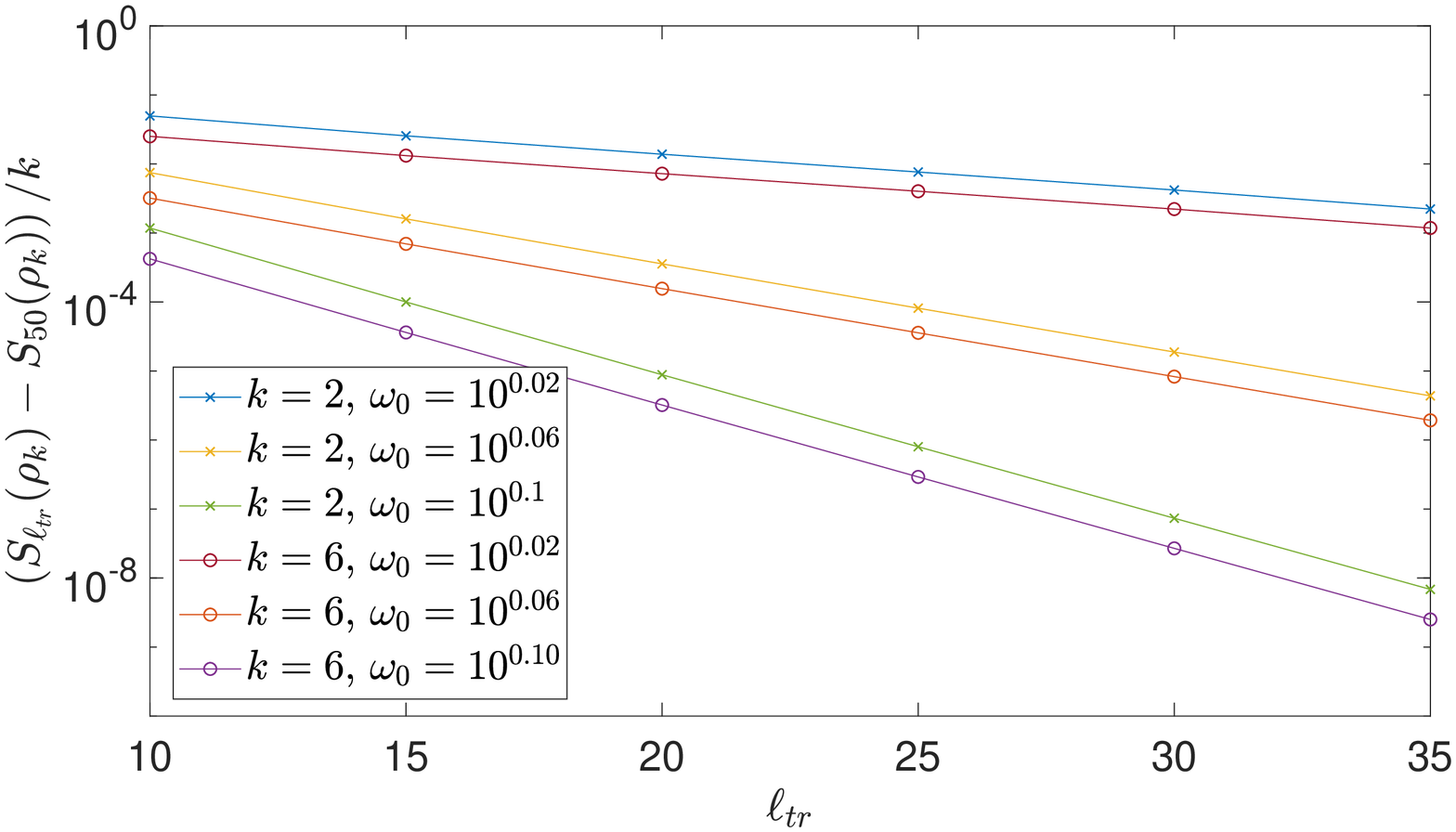}
     \includegraphics[scale=0.3]{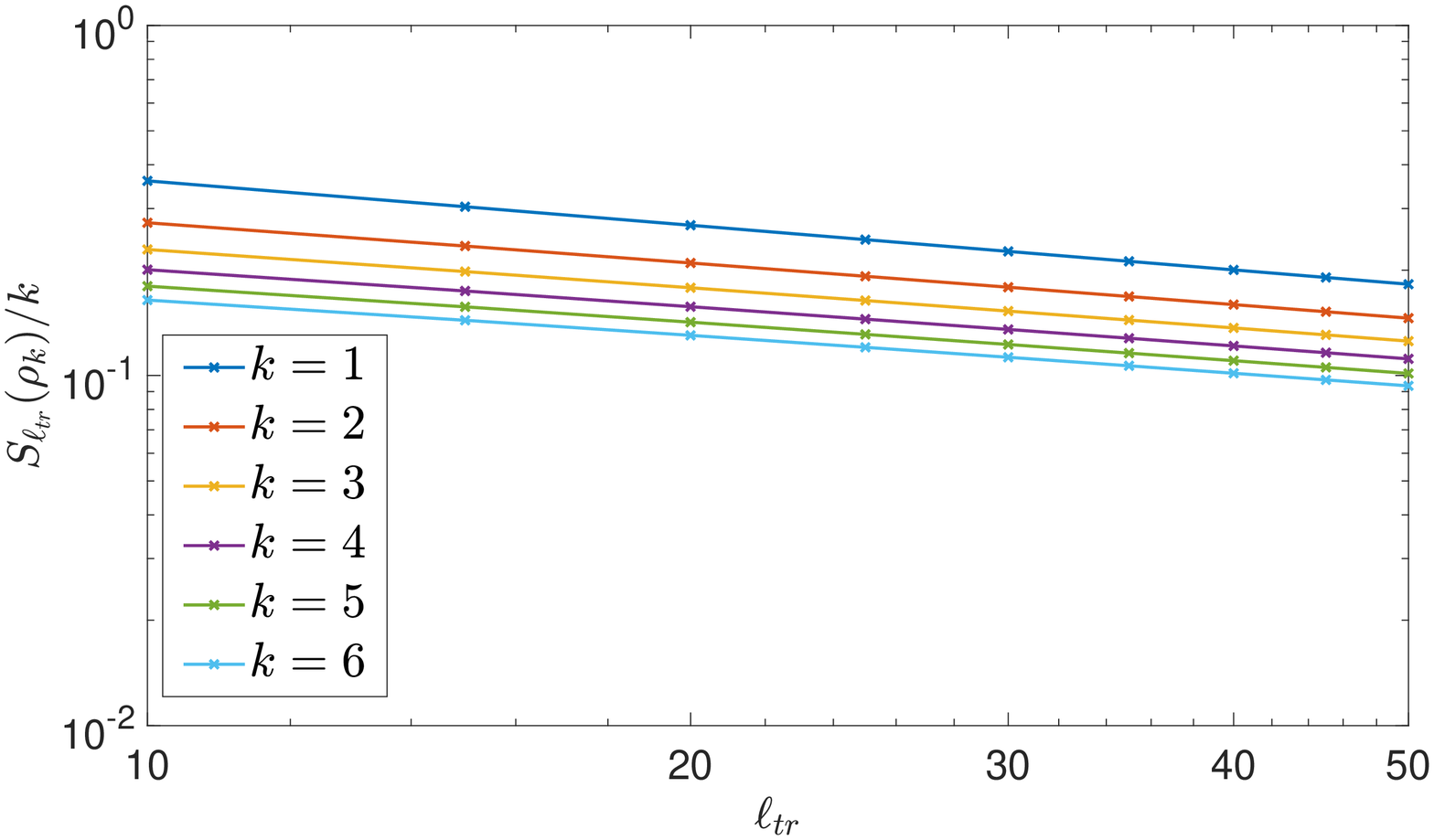}
    \caption{ Panel on the left on log-lin scale shows the truncation approaching to the exact NESS exponentially fast for $\omega_0 > 1$ and $k=2,6$, while the one in the right on log-log scale shows the truncation approaching in a power law behavior to the exact NESS for $\omega_0 = 1$ and $k=1,\dots,6$.}
    \label{fig:GMC_var_ltruncation}
\end{figure}
    We see that $S_{\ell_{tr}}(\rho_k)/k$ approaches exponentially fast to the NESS with varying truncation thresholds on the BTC phase ($\omega_0>1$), see left panel, while this quantity follows a power law for the critical coupling $\omega_0=1$ (right panel). % see Fig.\eqref{fig:GMC_var_ltruncation}. 
    For practical purposes, the NESS was considered with $50$ spins. More specifically,

\begin{eqnarray}
 S_{\ell_{tr}}(\hat{\rho}_k)/k \sim A_{\omega_0} e^{-\alpha_{\omega_0} \ell_{tr}} + B_{\omega_0},\qquad \omega_0 > 1,\\
 S_{\ell_{tr}}(\hat{\rho}_k)/k \sim A_{\omega_0} \ell_{tr}^{-\alpha_{\omega_0}} + B_{\omega_0},\qquad \omega_0 = 1,
\end{eqnarray}
where $B_{\omega_0} \equiv S_{\ell_{tr}}(\hat{\rho}_{ss})/k$, and $A_{\omega_0}$ and $\alpha_{\omega_0}$ are constants.

Extrapolating the above scaling, we can obtain the exact $S_{\ell_{tr}}(\hat{\rho}_k)/k$ for the exact NESS ($B_{\omega_0}$ in the above notation), as shown in Fig. \ref{fig:GMC_var_omega_ltr}. We see that for $\ell_{tr}=10$, with $k=1,2,6$, it approaches to the exact NESS for $\omega_0\sim 1.2$. Increasing the values of $\ell_{tr}$, for instance $\ell_{tr}=50$, we have $\omega_0\sim 1.05$, so that in the limit $S_{\ell_{tr \rightarrow \infty}}(\hat{\rho}_k)/k \rightarrow 0$ for $\omega_0 \rightarrow 1$. This behavior is in agreement with the GMC's presented in Fig. \ref{fig:NESSvshx}(d).
\begin{figure}[ht]
    \includegraphics[scale=0.3]{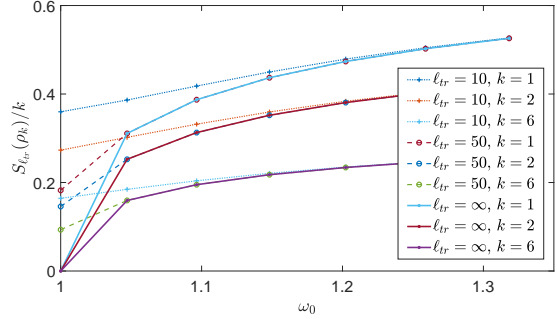}
    \caption{$S_{\ell_{tr}}(\hat{\rho}_k)/k$ as function of the pumping frequency $\omega_0$ for different partition sizes $k=1,2,6$ and number of spins in the system. As $\ell_{tr}$ increases the system state approaches the exact NESS.}
\label{fig:GMC_var_omega_ltr}
\end{figure}

   \begin{comment}
       
     \begin{equation}
        \hat{\rho}_{ss}=\frac{1}{D}\hat{\eta}\hat{\eta}^{\dagger}=\frac{1}{D}\sum_{j,l=0}^{\infty}\frac{(\hat{S}^-)^j(\hat{S}^+)^l}{(g^*)^jg^l},
    \end{equation}
    
    \end{comment}
    
    \subsection{Computation of the NESS observables}
    In order to compute the $\hat{\rho}_{ss}$ observables, we begin calculating the trace $\Tr{S^{-j}S^{+j}}$. Notice first that
    
    \begin{equation}
        S^{-j}S^{+j}\ket{S,s_z}= \prod_{s=s_z}^{s_z+j-1}\left[ S(S+1) - s(s+1) \right]\ket{S,s_z}.
    \end{equation}
    with $S=N/2$. Thus,
    \begin{align}
         \Tr{S^{-j}S^{+j}}=\sum_{s_z=-S}^{+S} \prod_{s=s_z}^{s_z+j-1} \left[ S(S+1) - s(s+1) \right]
          \xrightarrow[\substack{\lim S \rightarrow \infty \\ \lim j/S \rightarrow 0} ]{} \sum_{s_z=-S}^{+S}(S^2 - s_z^2)^j, 
    \end{align}
    results in
    \begin{align}
        \Tr{S^{-j}S^{+j}}=\sum_{s_z=-S}^{+S} \sum_{m=0}^{j}\binom{j}{m} (S^2)^{j-m}(-s_z^2)^m
        =\sum_{m=0}^{j}\binom{j}{m} (S^2)^{j-m}(-1)^m\left[\sum_{s_z=-S}^{+S}(s_z^2)^m\right].
    \end{align}
    In the case in which $\lim S \rightarrow \infty$ the Faulhaber's formula can be used, such that $$2\sum_{s_z=-S}^{+S}(s_z^2)^m\simeq \frac{2S^{2m+1}}{2m+1},$$
    then
    \begin{equation}
       \lim_{\substack{S \rightarrow \infty \\ j/S \rightarrow 0}}\Tr{S^{-j}S^{+j}}=2S^{2j+1}\sum_{m=0}^{j}\frac{(-1)^m}{2m+1}\binom{j}{m}. 
    \end{equation}
    One can see that, within these limits ($S\rightarrow \infty, j/S \rightarrow 0$) it is also true that:
    \begin{align}
        \Tr{S^{-j_1}S^{+l_1}S^{-j_2}S^{+l_2}...S^{-j_n}S^{+l_n}}= 
        \delta_{\sum_i j_i=J,\sum_i l_i}\Tr{S^{-J}S^{+J}},
    \end{align}
    which was already computed above. Similarly, with the same reasoning as above, we see that
    %\begin{widetext}
        \begin{align}
            \Tr{S^{-j}S^{+j}S_z^l} \xrightarrow[j/S,l/S \rightarrow 0]{S\rightarrow \infty}  \sum_{m=0}^{j}\binom{j}{m} (S^2)^{j-m}(-s_z^2)^m &=\sum_{m=0}^{j}(-1)^m\binom{j}{m} (S^2)^{j-m}\sum_{s_z=-S}^{+S}(s_z)^{2m+l} \nonumber \\
            &=\delta_{\mod_2 l,0} 2S^{2j+l+1}\sum_{m=0}^{j}\frac{(-1)^m}{2m+l+1}\binom{j}{m}.
        \end{align}
    %\end{widetext}
    Consequently, we have
    \begin{align}
        D=\Tr{\hat{\rho}_{ss}}=\sum_{j=0}^{\infty}\frac{\Tr{S^{-j}S^{+j}}}{|g|^{2j}}  =2\sum_{j=0}^{\infty}\frac{S^{2j+1}}{|g|^{2j}}\sum_{m=0}^{j}\frac{(-1)^m}{2m+1}\binom{j}{m}
        =2S\sum_{j=0}^{\infty}\frac{1}{|\tilde{g}|^{2j}}\sum_{m=0}^{j}\frac{(-1)^m}{2m+1}\binom{j}{m},
    \end{align}
    with $\tilde{g}=g/S$ and $g=(iwS)/k$. Also
    \begin{align}
        D\langle S^{-j_1}S^{+j_2}S_{z}^{j_3} \rangle = \sum_{j,l=0}^{\infty}\frac{\Tr{S^{-(j_1+j)}S^{+(j_2+l)}S_z^{j_3}}}{g^{*j}g^{l}} 
        =\sum_{\substack{j'=j_1 \\ l'=j_2}}^{\infty}\frac{\Tr{S^{-j'}S^{+l'}S_z^{j_3}}}{g^{*j'-j_1}g^{l'-j^2}},
    \end{align}
    we know $\delta_{J=j',l'}$, then
    %\begin{widetext}
        \begin{align}
            D\langle S^{-j_1}S^{+j_2}S_{z}^{j_3} \rangle &= \sum_{J=max(j_1,j_2)}^{\infty}\frac{\Tr{S^{-J}S^{+J}S_z^{j_3}}}{g^{*J-j_1}g^{J-j_2}} =\sum_{J=max(j_1,j_2)}^{\infty}\delta_{mod_2 j_3,0}\frac{2S^{2J+1+j_3}}{|g|^{2J}g^{*-j_1}g^{-j_2}}\sum_{m=0}^{J}\frac{(-1)^m}{2m+1+j_3}\binom{J}{m} \nonumber \\
            &=\delta_{mod_2 j_3,0}\frac{2S^{j_1+j_2+j_3+1}}{\tilde{g}^{*-j_1}\tilde{g}^{-j_2}}\sum_{J=max(j_1,j_2)}^{\infty}\frac{1}{|\tilde{g}|^{2J}}\sum_{m=0}^{J}\frac{(-1)^m}{2m+1+j_3}\binom{J}{m}.
        \end{align}
    %\end{widetext}
  
    \noindent Notice that the macroscopic observable $\hat{m}_\alpha\equiv \frac{\hat{S}_\alpha}{S}$, according to the above equations will be independent of "$S$", implying that
    \begin{equation}
        \langle \hat{m}_\alpha \hat{m}_\beta \rangle= \frac{1}{D}\frac{\Tr{\hat{S}_\alpha\hat{S}_\beta\rho_{ss}}}{S^2}=F(\alpha,\beta,\tilde{g}),
    \end{equation}
	in which the terms $O(S^3)$ have been neglected.
	%%%%%%%%%%%%%%%%%%%%%%%%%%%%%%%%%%%%%%%
	\subsection{Reduced density matrices (from collective spin observable-tomography)}
	%%%%%%%%%%%%%%%%%%%%%%%%%%%%%%%%%%%%%%%
	Based on the idea of quantum state tomography, a reduced state with $k$ particles out of $N$ can be written as
	\begin{equation}
	    \hat{\rho}_k=\Tr_{N-k}\{\hat{\rho}\}=\sum_{\alpha_1...\alpha_k=x,y,z,\mathbb{I}}\frac{\langle \hat{\sigma}_1^{\alpha_1}...\hat{\sigma}_k^{\alpha_k}\rangle}{2^k}\hat{\sigma}_1^{\alpha_1}...\hat{\sigma}_k^{\alpha_k}.
	\end{equation}
	
	We can infer the correlators $\langle \sigma_1 ... \sigma_k \rangle$ from the collective ones $\langle S_1 ... S_k \rangle$, as follows:
	%\begin{widetext}
	\begin{align}
	    \frac{1}{S^{k}}\langle \hat{S}^{\alpha_1}...\hat{S}^{\alpha_k}\rangle=\frac{1}{S^k}\sum_{i_1...i_k=1}^N \langle\hat{\sigma}_{i_1}^{\alpha_1}...\hat{\sigma}_{i_k}^{\alpha_k}\rangle\frac{1}{2^k} 
	    =\frac{1}{(2S)^k}\left[\sum_{\substack{\{i_k\} \\ i_1 \neq i_2...\neq i_k}}\langle \hat{\sigma}_{i_1}^{\alpha_1}...\hat{\sigma}_{i_k}^{\alpha_k}\rangle+\sum_{\substack{\{i_k\}\\ i_1 \neq i_{j\geq 2}}}\langle\hat{\sigma}_{i_1}^{\alpha_1}...\hat{\sigma}_{i_k}^{\alpha_k}\rangle+...\right].
	\end{align}
	%\end{widetext}
		\begin{figure}
        \includegraphics[scale=0.6]{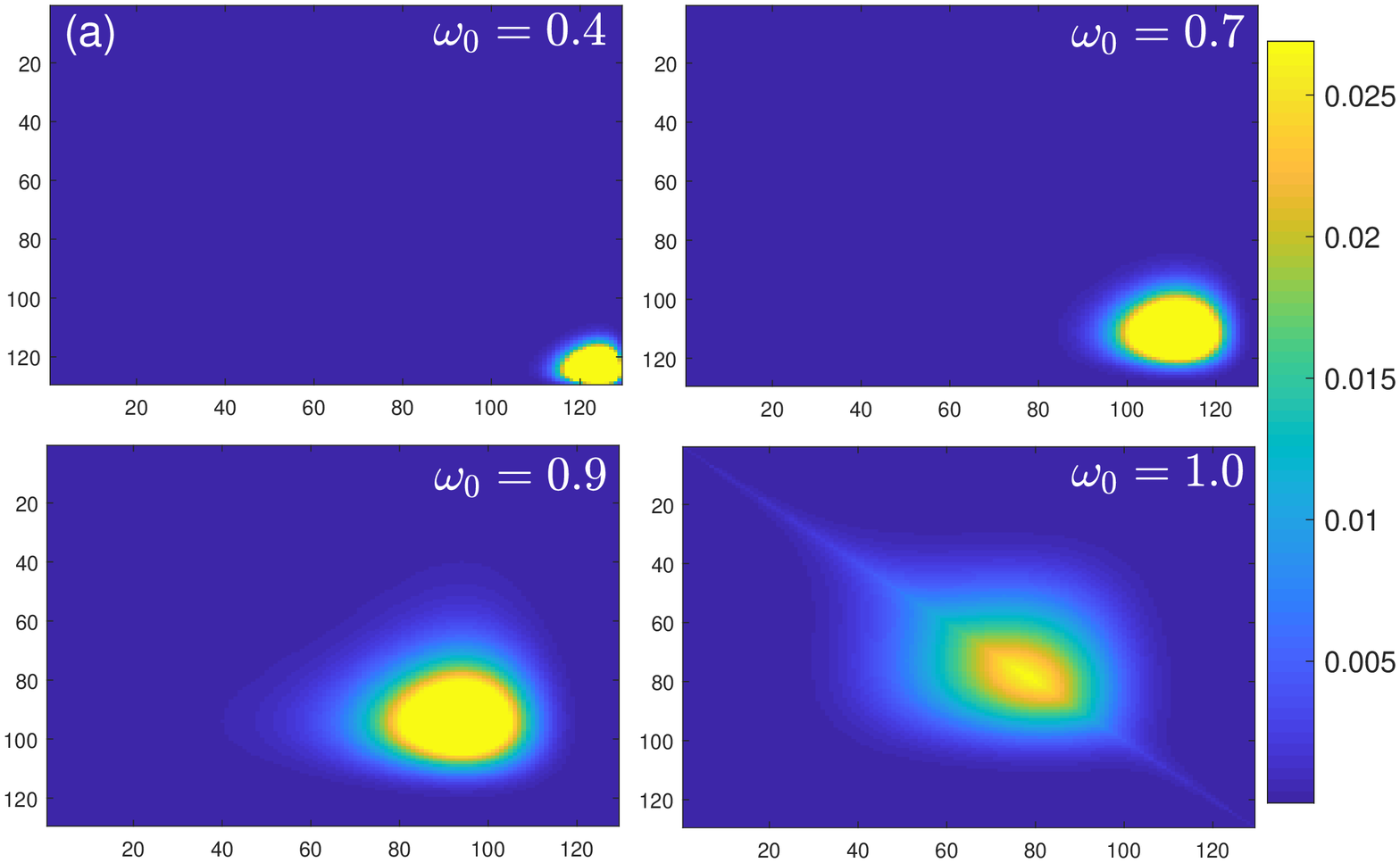}
        \includegraphics[scale=0.6]{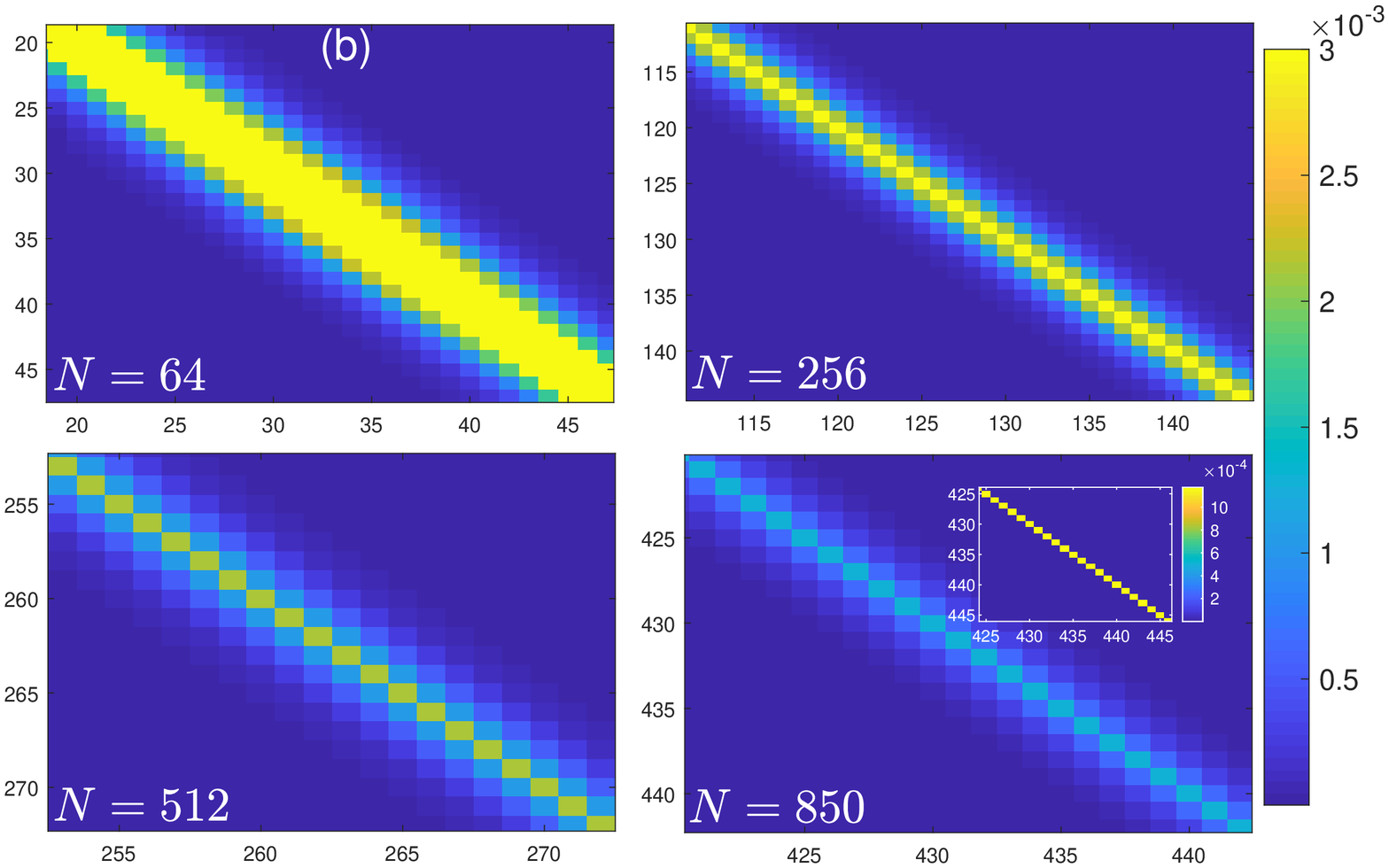}
    	\caption{Absolute values of the density matrix elements in a colormap in Dicke basis  for the NESS of the system along the phases of the model. The first element of the row and column represents all spins in excited state, while the last one all spins are in the ground state. We show in panel \textbf{(a)} for $N=128$ the behavior of density matrix in the ferromagnetic phase when increasing $\omega_0$ until its critical value $(\omega_0)_c = 1$. Panel \textbf{(b)} shows the zoom in the diagonal of the density matrix in the BTC phase with $\omega_0=2.0$ for some values of $N$. In the inset of right lower panel the absolute values of the density matrix for $N=850$ and $\omega_0=500.0$.
    	%\sout{When the system approaches the thermodynamic limit the matrix seems to become completely diagonal.}}  
    	}
    	\label{fig:NESSrho}
     \end{figure}
	Due to the particles permutation symmetry the expected value is independent of site indexes, i.e., particle indexes, then
	\begin{align}
	    \frac{1}{S^{k}}\langle \hat{S}^{\alpha_1}...\hat{S}^{\alpha_k}\rangle=\frac{1}{S^k}\sum_{i_1...i_k=1}^N \langle\hat{\sigma}_{i_1}^{\alpha_1}...\hat{\sigma}_{i_k}^{\alpha_k}\rangle\frac{1}{2^k} 
	    =\frac{1}{(2S)^k}[\langle \hat{\sigma}_{i_1}^{\alpha_1}...\hat{\sigma}_{i_k}^{\alpha_k}\rangle_{ i_1 \neq i_2...\neq i_k}\sum_{\substack{\{i_k\} \\ i_1 \neq i_2...\neq i_k}} 
	     +\langle\hat{\sigma}_{i_1}^{\alpha_1}...\hat{\sigma}_{i_k}^{\alpha_k}\rangle_{ i_1 \neq i_3...\neq i_k}\sum_{\substack{\{i_k\}\\ i_1 \neq i_{j\geq 2}}}+...],
	\end{align}
	where in the limit $S\rightarrow \infty$ the first sum becomes $\binom{N}{k}\sim \frac{N^k}{k!}\sim O(N^K)$ and the second one $\sim \binom{N}{k-1}\sim O(N^{k-1})$, with $(2S)^k=N^k$. Taking only the leading term $\langle \hat{\sigma}_{i_1}^{\alpha_1}...\hat{\sigma}_{i_k}^{\alpha_k}\rangle_{ i_1 \neq i_2...\neq i_k}/k!$ we get
	\begin{equation}
	    \langle \hat{\sigma}_1^{\alpha_1}...\hat{\sigma}_k^{\alpha_k}\rangle=k!\frac{\langle \hat{S}^{\alpha_1}...\hat{S}^{\alpha_k}\rangle}{S^{k}}.
	\end{equation}
	Therefore, the reduced state can be described as,
	\begin{equation}
	  \lim_{\substack{ S \rightarrow \infty \\ k/S \rightarrow 0}} \hat{\rho}_k = \sum_{\alpha_1...\alpha_k=x,y,z,\mathbb{I}}k!\frac{\langle\hat{S}^{\alpha_1}...\hat{S}^{\alpha_k}\rangle}{(2S)^k}\hat{\sigma}_1^{\alpha_1}...\hat{\sigma}_k^{\alpha_k}.  
	\end{equation}
	
	 %  \begin{figure}
     %   \includegraphics[scale=0.8]{NESS_BTC_N850_w10.eps}
    %	\caption{\textcolor{red}{Density matrix elements in a colormap in Dicke basis  for the NESS of the system in the BTC phase. In the upper panel is shown the absolute value of the matrix for N=850 and $\omega_0=10.0$, the matrix is almost diagonal and the terms of the diagonal comes from real part and off diagonal terms come from imaginary part of the elements of matrix, as shown in lowers panels. In the real and imaginary part images was taken absolute values of matrix to make the colormaps in similar colors.}  
    %	}
    %	\label{fig:NESSrhow10}
  %\end{figure}
        
	 \section{Coherence in the BTC} \label{app:coherence}
	 
	 In order to further discuss the qualitative behavior of the GMC's and QFI we explore the role played by quantum coherence \cite{baumgratz2014quantifying} in the Dicke basis in the NESS and in the system dynamics for finite times.

	 \subsection{Coherence of the NESS}
	
	 The Fig \ref{fig:NESSrho}(a) shows for $N=128$ that more close the system is to the phase transition $\omega_0 = (\omega_0)_c = 1$, more coherence the density matrix of total system has. Although the existence of quantum coherence in a system does not imply it is entangled, once coherence is basis dependent, it may work as resource for entanglement \cite{streltsov2017colloquium}. This  indicate that in the ferromagnetic phase the system is entangled and intensify this correlations close to the quantum phase transition, as shown in Fig. \ref{fig:FmaxNESSvshx}. 
	 
	 	\begin{figure}
        \includegraphics[scale=0.6]{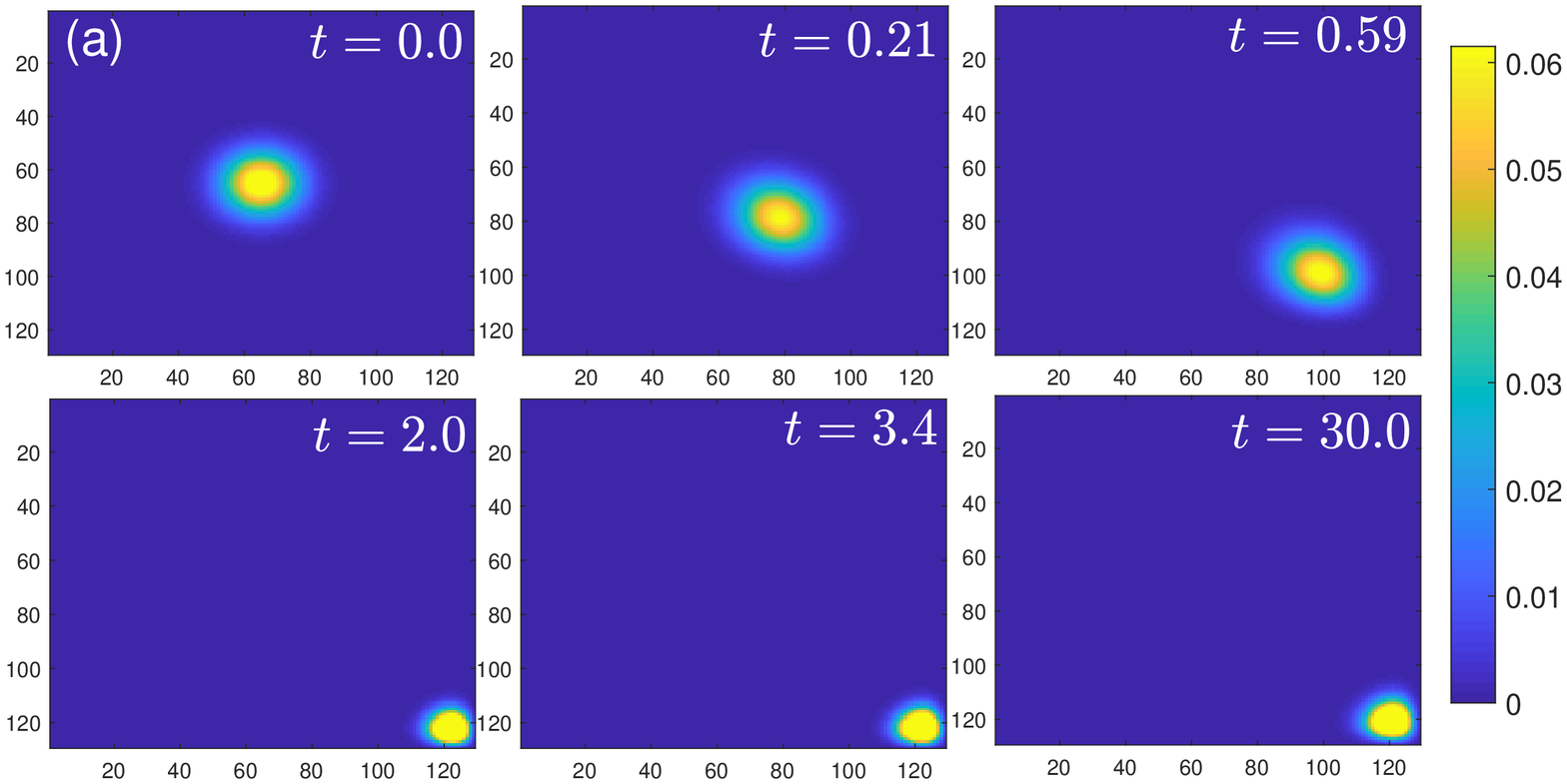}
        \includegraphics[scale=0.6]{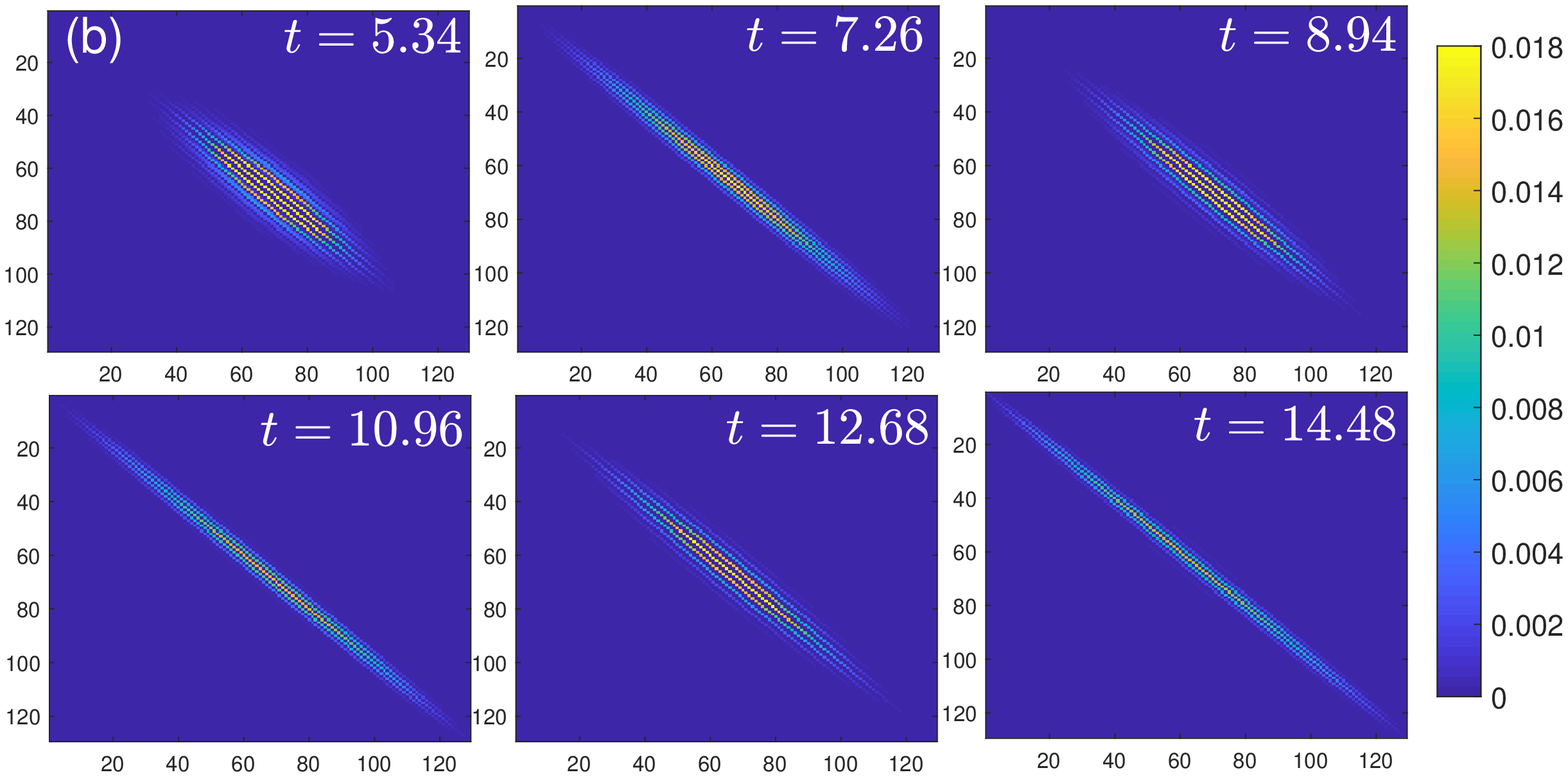}
    	\caption{Absolute values of the density matrix elements in a colormap in Dicke basis with $N=128$ for the dynamics of the system along the two phases of the model. The first element of the row and column means all spins in excited state and the last one all spins are in ground state. We show in panel \textbf{(a)}  the density matrix in the ferromagnetic phase ($\omega_0=0.5$) for some times of the dynamics, starting from a pure separable state $|\psi(0) \rangle = |-\rangle ^{\otimes N}$ at $t=0$ until reaching the NESS at $t=30.0$. Panel \textbf{(b)} shows the density matrix of the system in the BTC phase with $\omega_0=2.0$. At the time $t=5.34$ we find a valley in the dynamics of GMC's ($I^k$), so that in the next time $t=7.26$ a peak in $I^k$ is found. This alternating behavior is maintained until $t=14.48$, which enable us to see the matrix approaching to the quasi-diagonal NESS.  
    	}
    	\label{fig:Dynamicsrho}
	    \end{figure} 
  
	 On the other hand, observing the density matrix of the system $\rho_N$ in the BTC phase for large $N$ in the NESS as shown in Fig. \ref{fig:NESSrho}(b), we can see that it approaches to a diagonal matrix \cite{piccitto2021}, or a linear convex combination of Dicke states, as happens in Dicke superradiance phenomenon \cite{dicke54}.  Due to the limitation of computational resources, we plot the density matrix elements until $N=850$.
	  Despite an almost diagonal form, the NESS still have a nonnull coherence in the thermodynamic limit for finite $\omega_0/\kappa$. Notwithstanding, we obtain that the adjacent elements of the main diagonal are decreasing in the extremal limit with $\omega_0/\kappa \rightarrow \infty$. This can be observed in the inset of the right lower panel of the Fig. \ref{fig:NESSrho}(b), which shows the absolute value of the matrix elements for $\omega_0/\kappa=500$. As demonstrated in Refs. \cite{wolfe2014,yu2016separability,tura2018separability} a state in this form is separable, then, pointing towards that the NESS of the system in this limit is not entangled. Despite of that, this a quantum state, once it has discord-like quantum correlations  \cite{santos2016}.

   \subsection{Coherence of the dynamical state}
   
   In Fig.~\ref{fig:Dynamicsrho}(a) we can see the dynamics of the system in the ferromagnetic phase with $\omega_0 =0.5$ and $N=128$ spins. The initial pure state $|\psi(0) \rangle = |-\rangle ^{\otimes N}$ at $t=0$ is separable and as time passes it is driven to Dicke states that have mostly the spins in the ground state, which are approximately described by few excitations of the state $|N,0 \rangle$. What prevents the NESS of being entirely in $\ket{N,0}$ is the weak field $\omega_0$. Here, we call the attention to the fact that the initial state has coherence in the Dicke basis, although it is separable. This result is in accordance with the discussion above, once the existence of quantum coherence does not imply in entanglement, but it can be converted to entanglement through incoherent operations \cite{streltsov2015measuring}.

   Fig.~\ref{fig:Dynamicsrho}(b) shows the dynamics of the system in the BTC phase for $\omega_0 =2.0$ and $N=128$. The times were chosen to capture the valleys and peaks of GMC's (as in Fig. \ref{fig:STL120w05}(b), for instance), so that at time $t=5.34$ it describes a valley and at the second time $t=7.26$ it is in a peak, and so on alternately. We notice that in the peaks of GMC's the density matrix is near its quasi-diagonal form, while in the valleys it has larger coherence. Similarly to NESS in the BTC phase, the states of the system in the peaks are almost diagonal and present higher values of GMC's compared to the ferromagnetic phase. %\sout{As explained before, it is likely that these are quantum discord-like correlations or classical ones.} }  

	\end{widetext}
	\end{appendix}
	
	\bibliography{references}

\end{document}